\documentclass[journal]{IEEEtran}

\ifCLASSINFOpdf
\else
   \usepackage[dvips]{graphicx}
\fi
\usepackage{url}

\usepackage{etoolbox}%
\robustify\bfseries
\usepackage{graphicx}
\usepackage{amsmath,amssymb,graphicx}
\usepackage[]{algorithm2e}
\usepackage{tabularx}
\usepackage{wasysym}
\usepackage{color,soul}
\usepackage{cite}
\usepackage{balance}
\usepackage{breqn}
\usepackage{xcolor}
\usepackage{makecell}
\usepackage{multirow}
\usepackage{siunitx}
\usepackage{booktabs}
\def\H{{\mathsf H}}
\def\T{{\mathsf T}}
\def\CC{{\mathbb C}}

\usepackage{soul}
\soulregister\cite7 %
\soulregister\citep7 %
\soulregister\citet7 %
\soulregister\ref7 %
\soulregister\pageref7 %
\newcommand{\ZQHL}[1]{#1} %
\newcommand{\ZQHLAQ}[1]{#1} %

\newcommand{\remove}[1]{}

\makeatletter
\newcommand*{\bigs}[1]{{\hbox{$\left#1\vbox to9\p@{}\right.\n@space$}}}
\makeatother

\begin{document}

\title{Convolutive Prediction\\for Monaural Speech Dereverberation\\and Noisy-Reverberant Speaker Separation}

\author{Zhong-Qiu Wang, Gordon Wichern, and Jonathan Le Roux
\thanks{Manuscript received on Jan. 25, 2021; revised Jul. 11, 2021; revised Oct. 12, 2021.}
\thanks{Z.-Q. Wang, G. Wichern, and J. Le Roux are with Mitsubishi Electric Research Laboratories, Cambridge, MA 02139, USA (e-mail: wang.zhongqiu41@gmail.com, \{wichern,leroux\}@merl.com).}}

\maketitle

\begin{abstract}
A promising approach for speech dereverberation is based on supervised learning, where a deep neural network (DNN) is trained to predict the direct sound from noisy-reverberant speech.
This data-driven approach is based on leveraging prior knowledge of clean speech patterns, and \ZQHLAQ{seldom} explicitly exploits the linear-filter structure in reverberation, i.e., that reverberation results from a linear convolution between a room impulse response (RIR) and a dry source signal.
In this work, we propose to exploit this linear-filter structure within a deep learning based monaural speech dereverberation framework.
The key idea is to first estimate the direct-path signal of the target speaker using a DNN and then identify signals that are decayed and delayed copies of the estimated direct-path signal, as these 
can be reliably considered as reverberation.
They can be either directly removed for dereverberation, or used as extra features for another DNN to perform better dereverberation.
To identify the copies, we estimate the underlying filter (or RIR) by efficiently solving a linear regression problem per frequency in the time-frequency domain.
We then modify the proposed algorithm for speaker separation in reverberant and noisy-reverberant conditions.
State-of-the-art speech dereverberation and speaker separation results are obtained on the REVERB, SMS-WSJ, and WHAMR! datasets.

\end{abstract}


\IEEEpeerreviewmaketitle

\section{Introduction}

\IEEEPARstart{R}{oom} reverberation is pervasive in modern hands-free speech communication such as teleconferencing and interaction with smart speakers.
In reverberant rooms, speech signals propagate in the air and are inevitably reflected by the walls, floor, ceiling, and any other objects in the room before being captured by far-field microphones. Reverberation can be considered as a summation of an infinite number of decayed and delayed copies of source signals.
It degrades speech intelligibility and quality and is harmful to modern automatic speech recognition (ASR) systems. Dereverberation is a challenging task, as it is hard to pinpoint the direct-path signal and differentiate it from its copies, especially when reverberation is strong and non-stationary noises are also present. Different from blind deconvolution, which is not solvable without making assumptions on input signals or impulse responses \cite{Levin2011, blinddeconvolution}, speech dereverberation is simpler, as in the time-frequency (T-F) domain speech exhibits unique patterns, which provide an informative cue for reverberation reduction.

The most popular dereverberation algorithm is weighted prediction error (WPE) \cite{Nakatani2010}.
It computes a filter based on variance-normalized delayed linear prediction to estimate late reverberation from past observations, which is then subtracted from the mixture to estimate target speech. It iteratively estimates the time-varying power spectral density (PSD) of the target speech and the linear filter in an unsupervised manner.
WPE is found to introduce little distortion to target speech and leads to consistent improvements in many robust ASR tasks\ZQHLAQ{\cite{Kinoshita2016, Barker2018CHiME5}}. 
Other conventional approaches for dereverberation include estimating a Wiener-like filter based on estimated reverberation time \cite{Habets2009}, on the estimated PSD of late reverberation \cite{Braun2018}, or on a relative convolutive transfer function model \cite{Talmon2009}.

Another popular approach is based on supervised learning, where DNNs are trained to directly predict target speech from reverberant speech \cite{WDL2018}.
This approach is flexible at dealing with many related tasks.
For example, it can use noisy-reverberant speech as input for noisy-reverberant speech enhancement or multi-speaker separation, depending on the targets and loss functions used for model training.
In monaural dereverberation, a DNN was initially used in the magnitude domain \cite{Han2015} to predict T-F masks or target magnitudes with or without additional magnitude-based phase reconstruction.
In the DNN-WPE algorithm \cite{Kinoshita2017}, DNN-estimated magnitudes are used to compute the target PSD in WPE so that WPE's iterative procedure is avoided.
Subsequently, DNNs have been utilized to estimate complex T-F domain \cite{Wang2020b, Wang2020a, Wang2020c} and time domain \cite{Luo2018, Luo2019, Luo2020} signals that model phase and magnitude at the same time.
Riding on the development of deep learning,
many recent studies along this line \cite{Luo2020, Wang2020b, Zhao2020, J.Borgstrom2020} focus on adapting novel neural network architectures such as dilated convolution, self-attention, and recurrent networks for more end-to-end modeling.
\ZQHL{Such deep learning based approaches \cite{WDL2018}, despite often using neural networks as a black box and not heavily relying on conventional signal processing algorithms, can model speech patterns very well.
They have been firmly established as the state-of-the-art technique in speech enhancement and speaker separation.
However, one largely missing part is that many speech dereverberation studies do not explicitly exploit the linear convolutional structure of reverberation.}
DNN-WPE \cite{Kinoshita2017} is one exception, but it suffers from some shortcomings.
\ZQHL{First, to avoid cancellation in target speech, WPE uses a prediction delay \cite{Nakatani2010}, which is likely to limit its capability at removing early reflections.
Second, DNN-WPE is designed to only leverage the estimated magnitude produced by DNNs \cite{Kinoshita2017}.
It lacks a mechanism to leverage DNN-estimated phase, which could be utilized to design a new form of linear prediction for better dereverberation.
Third, as competing speakers or noises become stronger, monaural DNN-WPE becomes gradually biased towards reducing the reverberation of these sources rather than that of the target source. We will give a detailed analysis of this biasing problem in Section~\ref{robustnessdescription}.}

In this context, our study investigates the combination of linear prediction and deep learning for monaural speech dereverberation, where we first use a DNN to estimate target anechoic speech and then find delayed and decayed copies of the estimated anechoic speech by solving a linear regression problem.
Such copies can be safely removed for dereverberation as they are repetitive patterns likely due to reverberation.
They can also be used as extra features to train another DNN for better dereverberation.
We extend this approach to monaural talker-independent multi-speaker separation in reverberant as well as noisy-reverberant conditions, where we find delayed and decayed copies of each speaker for dereverberation.

We make two major contributions on monaural dereverberation.
First, we propose convolutive prediction, a novel formulation of linear prediction that can utilize DNN-provided target statistics for speech dereverberation.
Compared with WPE and DNN-WPE \cite{Kinoshita2017}, convolutive prediction, \ZQHLAQ{in the forward-filtering case,} can better reduce early reflections, can use estimates of both target magnitude and phase obtained by DNNs for linear prediction, and can better estimate a dereverberation filter for each source when there are multiple target sources.
Second, we use convolutive prediction outputs to train another DNN for better dereverberation. We find that such outputs contain complementary information to deep learning based end-to-end approaches.
As a more minor contribution, we propose a new loss function that can improve \ZQHL{typical permutation-free objective functions (also known as permutation invariant training (PIT)) \cite{R.Hershey2016,Isik2016,Kolbak2017} for noisy-reverberant speaker separation.}
The proposed system obtains state-of-the-art performance on three datasets including REVERB \cite{Kinoshita2016}, SMS-WSJ \cite{Drude2019}, and WHAMR! \cite{Maciejewski2020}.

In the rest of this paper, we describe the hypothesized physical model in Section \ref{phymodel}, give an overview of our system in Section \ref{systemoverview}, review WPE and DNN-WPE in Section \ref{wpereview}, and detail the proposed convolutive prediction in Section \ref{cpdescription} and DNN configurations in Section \ref{dnndescription}. Experimental setup is presented in Section \ref{setup}, followed by evaluation results in Section \ref{results} and conclusions in Section \ref{conclusion}.

\section{Physical Model and Objectives}\label{phymodel}
Given a monaural signal recorded in a noisy-reverberant setting, the physical model describing the relationships between the mixture $y$, reverberant target speech $x$, and non-target sources $v$ including reverberant noises and reverberant competing speakers can be formulated in the time domain as
\begin{align} 
	y[n] &= x[n]+v[n] = (a*r)[n]+v[n] \nonumber \\
	&= (a*r_d)[n] + (a*r_e+a*r_l)[n] + v[n] \nonumber \\
	&= s[n] + h[n] + v[n], \label{eq:phymodel_time}
\end{align}
where %
$n$ indexes discrete time, $*$ denotes the convolution operator, and $x$ results from a linear convolution between a dry source signal $a$ and an RIR $r$.
\ZQHL{We use $r_d$, $r_e$, and $r_l$ to respectively denote the direct, early, and late parts of the RIR.
The direct-path signal (or direct sound) $s$ is defined as $s=a*r_d$, while the non-direct-path signal $h$ is defined as the summation of the early reflections $a*r_e$ and late reverberation $a*r_l$, i.e., $h=a*r_e+a*r_l$.
Note that we define impulses up to 50 ms \cite{Yoshioka2012} after the direct-path peak of $r$ as $r_{d+e}$, and $r_e$ as $r_e=r_{d+e}-r_d$.}
In the short-time Fourier transform (STFT) domain, the physical model is formulated as
\begin{align} 
	Y(t,f) &= X(t,f)+V(t,f) \nonumber \\
	&= S(t,f)+H(t,f)+V(t,f), \label{eq:phymodel_freq}
\end{align}
where $Y(t,f)$, $X(t,f)$, $S(t,f)$, $H(t,f)$, and $V(t,f)$ respectively denote the STFT coefficients of the captured mixture, reverberant target speech, direct-path signal, early reflections plus late reverberation, and non-target sources at time $t$ and frequency $f$. The corresponding spectrograms are denoted by $Y$, $X$, $S$, $H$, and $V$.

We aim at recovering $S$ from $Y$, and use $s$ as the reference signal for metric computation. We emphasize that early reflections are not considered as part of target speech.

Our study considers three tasks: speech dereverberation, reverberant speaker separation, and noisy-reverberant speaker separation. These three tasks are progressively more difficult, as we include competing speakers and non-stationary noises, which are known to be very detrimental to linear prediction.

Since this work mainly focuses on suppressing reverberation, most equations in this paper are  designed for speech dereverberation, where we assume that there is only one speaker active.
For multi-speaker separation, we do not explicitly write out multiple speakers in the above formulations as well as in later proposed algorithms, in order to make equations less cluttered.
\ZQHL{However, readers should be aware that for the convolutive prediction in multi-speaker scenarios, we consider each speaker as the target speaker in turn, with the remaining speakers considered as competing speakers and absorbed into $v$.}
We will introduce a speaker index in our notations when dealing with speaker separation in Section \ref{dnndescription}.

\begin{figure}
  \centering  
  \includegraphics[width=9cm]{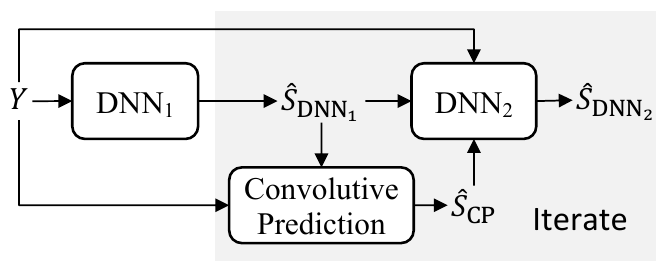} \vspace{-0.7cm} \\  
  \caption{\footnotesize{Illustration of the proposed speech dereverberation system.}}
  \label{systemenh}\vspace{-0.4cm}
\end{figure} 

\section{System Overview}\label{systemoverview}

Fig.~\ref{systemenh} illustrates our system for speech dereverberation. It contains two DNNs and a linear-prediction module in between. 
The first DNN estimates target anechoic speech, which is then used to estimate the reverberation of the target speaker based on convolutive prediction.
We then combine the mixture and the outputs from the first DNN and from convolutive prediction
as inputs for a second DNN to further estimate target anechoic speech.
The output from the second DNN is expected to be better than that from the first one, so we can use it to do convolutive prediction again and run the second DNN for one more time.
This process can be iterated to gradually refine target estimation. 

Both networks are trained using complex spectral mapping, where we predict the real and imaginary (RI) components of target speech from the mixture RI components \cite{Wang2020b, Wang2020a, Wang2020c}.
For speech dereverberation, we define the loss based on the predicted RI components and target RI components.
For speaker separation, we additionally use PIT \cite{R.Hershey2016,Isik2016,Kolbak2017} to resolve the label-permutation problem.
More details on the DNNs will be provided in Section \ref{dnndescription}.
At this point, readers can assume that each DNN in our system can provide an estimate of target anechoic speech in the complex T-F domain, denoted as $\hat{S}_{\text{DNN}_b}$, where $b\in\{1,2\}$ as our system has two DNNs.

There are multiple options for the convolutive- or linear-prediction module. In Section \ref{wpereview}, we review WPE and DNN-WPE, which are based on linear prediction and considered as the most popular dereverberation algorithms to date. We propose convolutive prediction in Section \ref{cpdescription}.

\section{WPE and DNN-WPE}\label{wpereview}
This section reviews WPE and DNN-WPE, and analyzes their strengths and weaknesses, which motivated the design of our algorithms for dereverberation.

WPE \cite{Nakatani2010} computes an inverse linear filter per frequency to 
estimate the late reverberation at the current frame from past noisy-reverberant observations.
The estimated late reverberation is then subtracted from the mixture for dereverberation.
Mathematically, the dereverberation result is obtained as
\begin{align}\label{wperesult}
\hat{S}_{\text{WPE}}(t,f)=Y(t,f)-\hat{\mathbf{g}}(f)^{\H}\widetilde{\mathbf{Y}}(t-\Delta,f),
\end{align}
where $\hat{\mathbf{g}}(f)\in \CC^{K}$ denotes a $K$-tap complex-valued filter, $\widetilde{\mathbf{Y}}(t,f)=[Y(t,f),Y(t-1,f),\dots,Y(t-K+1,f)]^\T$, and $\Delta$ ($\geq 1$) is a  prediction delay.
Assuming that the estimated target speech follows a complex Gaussian distribution with zero mean and a time-varying PSD $\lambda(t,f)$, i.e., $S_{\text{WPE}}(t,f) \sim \mathcal{N}\big(0, \lambda(t,f)\big)$, and based on maximum likelihood estimation, WPE estimates the filter through the minimization problem
\begin{align}\label{wpe}
\underset{\mathbf{g}(f), \lambda(\cdot, f)}{{\text{argmin}}} \sum_{t} \frac{|Y(t,f)-\mathbf{g}(f)^{\H}\ \widetilde{\mathbf{Y}}(t-\Delta,f)|^2}{\lambda(t, f)}+\log %
\lambda(t, f),
\end{align}
where $|\cdot|$ computes magnitude. Eq.~(\ref{wpe}) does not have closed-form solutions. An iterative algorithm is proposed in \cite{Nakatani2010} to alternately estimate $\mathbf{g}(f)$ and $\lambda(t,f)$.

Note that $\Delta$ cannot be set to zero, because otherwise a trivial but useless solution $\hat{\mathbf{g}}(f)=[1,0,\dots,0]^{\T}$ reaches the global optimum.
Given a typical 32 ms STFT window size and an 8 ms hop size, \ZQHL{$\Delta$ is usually set by default, or tuned through a validation set, to 3 or 4.}
This is possibly because smaller $\Delta$ makes $Y(t,f)$ and $\widetilde{\mathbf{Y}}(t-\Delta,f)$ share more time-domain signal samples due to the overlap between nearby frames, and will thus more likely result in target speech cancellation.
However, a larger delay, for example set to a value greater than 3, will likely limit WPE's capability at reducing early reflections.
\ZQHLAQ{Note that early reflections can smear spectral patterns, albeit not as severely as late reverberation, and they are found to degrade ASR performance in \cite{Wang2020c}.}
Our proposed convolutive prediction algorithm aims at removing both early reflections and late reverberation. 

In the subsequent DNN-WPE algorithm \cite{Kinoshita2017}, $\lambda$ is no longer jointly estimated \ZQHL{via WPE's iterative optimization procedure}, but replaced by an estimate $\hat{\lambda}$ obtained by a DNN, and Eq.~(\ref{wpe}) is simplified to
\begin{align}\label{dnnwpe}
\underset{\mathbf{g}(f)}{{\text{argmin}}} \sum_{t} \frac{|Y(t,f)-\mathbf{g}(f)^{\H}\ \widetilde{\mathbf{Y}}(t-\Delta,f)|^2}{\hat{\lambda}(t, f)},
\end{align}
\ZQHL{which has a closed-form solution.}
The dereverberation result $\hat{S}_{\text{DNN-WPE}}$ is computed in the same way as in Eq.~(\ref{wperesult}).

In WPE, $\lambda$ is initialized based on mixture energy, and then iteratively updated.
In DNN-WPE, there are multiple options for $\lambda$. Depending on the DNN training target, $\lambda$ can be the estimated PSD of (a) the summation of the direct sound, early reflections, and noise, following \cite{Kinoshita2017}; (b) the direct sound plus early reflections, following \cite{Heymann2019, Haeb-Umbach2020}; or (c) the direct sound only.
\ZQHL{Let us denote by $\hat{Z}$ the magnitude spectrogram computed from DNN outputs}\footnote{\ZQHL{Our DNNs estimate the complex spectrogram of the target during training. We then compute $\hat{Z}$ based on the complex-domain estimate.}}.
For all three options, $\hat{\lambda}$ is computed as
\begin{align}\label{wpelambda}
\hat{\lambda}(t,f)=\text{max}\big(\varepsilon \text{max}(\hat{Z}^2),\hat{Z}(t,f)^2\big),
\end{align}
where $\text{max}(\cdot)$ extracts the maximum value of a spectrogram, $\text{max}(\cdot,\cdot)$ returns the larger of two values, and $\varepsilon$ is a floor value to avoid putting too much weight on silent T-F units. Setting $\varepsilon$ to one essentially means no weighting is used.
We will compare these options in our experiments.

DNN-WPE cleverly utilizes target statistics estimated by a DNN to avoid iterative optimization, paving the way towards online dereverberation \cite{Heymann2018b} and joint training of WPE with other DNN modules \cite{Heymann2019, Zhang2020}. Compared with WPE, which is unsupervised in nature, DNN-WPE can leverage the modeling power of DNN on magnitude-domain speech patterns to improve PSD estimation. Motivated by and building upon DNN-WPE, we explore other ways of using statistics provided by a DNN for linear prediction.

Our first insight for potential improvement is that DNN-WPE only utilizes the estimated target magnitude produced by the DNN (i.e., by using it to compute $\hat{\lambda}$), partly because only magnitude could be estimated reasonably well at that time, but not phase.
Recent deep learning studies have shown that phase estimation can also be improved by using deep learning \cite{Williamson2016, Wang2018d, Luo2019, Wang2020b, Zeghidour2020, Luo2020}.
\ZQHLAQ{An interesting question is whether we can design a new linear prediction algorithm that can leverage both magnitude and phase estimated by a DNN for dereverberation, and whether the new linear-prediction result can have less reverberation, or can be utilized as a better feature than the WPE result for the second DNN in Fig.~\ref{systemenh}.}

Another insight for potential improvement is that the linear filtering in WPE is applied to the mixture.
This means that WPE essentially estimates a single filter to reduce the reverberation of all sources.
Indeed, in the literature, WPE and DNN-WPE are typically used as a pre-processing before later enhancement/separation algorithms \cite{Boeddecker2018, Zhang2020}.
The rationale is that if the input mixture is overall less reverberant, later processing becomes easier.
However, \ZQHL{in the monaural case which we consider in this study}, computing a single filter to dereverberate the mixture is not good enough when noise or competing speakers are very strong, because the filter would be biased towards suppressing the reverberation of higher-energy sources \ZQHL{(see our discussion later in Section~\ref{robustnessdescription} and Eq.~(\ref{wperobust}) for detailed analysis).
We point out that in multi-microphone scenarios, a single multi-microphone input multi-microphone output filter could theoretically dereverberate all the sources \cite{Masato1988, Haeb-Umbach2020}, but one key constraint is that the number of microphones should be no fewer than the number of sources.
However, when there is only one microphone but multiple sources, this constraint is not satisfied.}
In such cases, it seems more promising %
to estimate a dereverberation filter for each source, as each source is convolved with a different RIR. In DNN-WPE, it is indeed possible to compute a different filter for each source, by using the estimated PSD of each source in turn in Eq.~(\ref{dnnwpe}). However, \ZQHLAQ{in the optimization problem the linear filter is estimated by multiplying it with the mixture}, which consists of multiple sources, and the estimated filter could still be biased towards dereverberating higher-energy sources.

We tackle these problems in the following section.

\section{Convolutive Prediction}\label{cpdescription}
\ZQHL{To address the aforementioned issues in DNN-WPE, we propose a new DNN-supported method to estimate a dereverberation filter which we refer to as forward convolutive prediction (FCP).
Based on a target speech estimate $\hat{S}_{\text{DNN}_b}$ produced by the DNN, FCP estimates a dereverberation filter by forwardly %
filtering $\hat{S}_{\text{DNN}_b}$ to approximate the mixture.
As a bridge between WPE and FCP, we also present inverse convolutive prediction (ICP), which estimates a dereverberation filter by inversely filtering the mixture to approximate $\hat{S}_{\text{DNN}_b}$.}
In Sections \ref{icpdescription} and \ref{fcpdescription}, we assume $V$ is weak, meaning that $Y\approx X$.
In Section \ref{robustnessdescription}, we will discuss the cases where $V$ is strong and non-negligible.

\subsection{Inverse Convolutive Prediction}\label{icpdescription}

In WPE, a linear time-invariant inverse filter is computed to approximate the late reverberation contained in $Y(t,f)$ based on the delayed mixture $\widetilde{\mathbf{Y}}(t-\Delta,f)$.
In ICP, we linearly filter $\widetilde{\mathbf{Y}}(t,f)$ to approximate $Y(t,f)-\hat{S}_{\text{DNN}_b}(t,f)$, which can be considered as the estimated reverberation if the interference $V(t,f)$ is weak.
\ZQHL{Similarly to our DNN-WPE setup, we compute $\hat{\lambda}(t,f)$ based on the target speech estimated by a DNN}, but we consider a slightly different minimization problem
\begin{align}\label{complexproj1dummy}
\underset{\mathbf{g}'(f)}{{\text{argmin}}} \sum_{t} \frac{|Y(t,f)-\hat{S}_{\text{DNN}_b}(t,f)-\mathbf{g}'(f)^{\H}\ \widetilde{\mathbf{Y}}(t,f)|^2}{\hat{\lambda}(t,f)},
\end{align}
and obtain the dereverberation result as $Y(t,f)-\hat{\mathbf{g}}'(f)^{\H}\widetilde{\mathbf{Y}}(t,f)$.
\ZQHL{As $Y(t,f)$ also appears in $\widetilde{\mathbf{Y}}(t,f)$, we can equivalently absorb $Y(t,f)$ into $\widetilde{\mathbf{Y}}(t,f)$ and reformulate the minimization problem of Eq.~(\ref{complexproj1dummy}) as}
\begin{align}\label{complexproj1}
\underset{\mathbf{g}(f)}{{\text{argmin}}} \sum_{t} \frac{|\hat{S}_{\text{DNN}_b}(t,f)-\mathbf{g}(f)^{\H}\ \widetilde{\mathbf{Y}}(t,f)|^2}{\hat{\lambda}(t,f)},
\end{align}
from which we obtain the dereverberation result as
\begin{align}
\hat{S}_{\text{ICP}}(t,f)=\hat{\mathbf{g}}(f)^{\H}\widetilde{\mathbf{Y}}(t,f).
\end{align}
\ZQHL{Note that we use a prime symbol to differentiate $\mathbf{g}'(f)$ in Eq.~}(\ref{complexproj1dummy}) with $\mathbf{g}(f)$ in (\ref{complexproj1}).
The estimated filter $\hat{\mathbf{g}}(f)$ acts as an inverse filter that seeks to undo the (forward) reverberation process in order to get from the reverberated mixture $Y$ back to the target direct-path speech $S$ (in lieu of the dry source signal, whose estimation is ill-defined).
The objective to minimize is quadratic, \ZQHLAQ{similar to that of DNN-WPE}, and a closed-form solution is readily available. Section \ref{lambda} will discuss how to set $\hat{\lambda}(t,f)$.

Eq.~(\ref{complexproj1}) essentially 
applies a time-invariant filter to the mixture to approximate the target speech estimated by the DNN.
This is in spirit similar to the classic multi-channel Wiener filter in microphone array processing \cite{Gannot2017}, where a linear filter per T-F unit is computed to 
estimate target speech from the multi-channel mixture mainly based on linear spatial cues.
The difference here is that we apply this strategy for monaural processing and to exploit the linear-filter structure in reverberation.

\subsection{Forward Convolutive Prediction}
\label{fcpdescription}

In FCP, we approximate 
$Y(t,f)-\hat{S}_{\text{DNN}_b}(t,f)$, which again can be considered as the estimated reverberation of the target speaker if we assume that $V(t,f)$ is weak, by forwardly filtering  the target speech estimate $\hat{S}_{\text{DNN}_b}$ obtained by the DNN. The filter is obtained by solving the problem
\begin{align}\label{complexproj2dummy}
\underset{\mathbf{g}'(f)}{{\text{argmin}}} \sum_{t} \frac{|Y(t,f)-\hat{S}_{\text{DNN}_b}(t,f)-\mathbf{g}'(f)^{\H}\ \widetilde{\hat{\mathbf{S}}}_{\text{DNN}_b}(t,f)|^2}{\hat{\lambda}(t,f)},
\end{align}
where $\widetilde{\hat{\mathbf{S}}}_{\text{DNN}_b}(t,f)\!=\![\hat{S}_{\text{DNN}_b}(t,f),\hat{S}_{\text{DNN}_b}(t\!-\!1,f),\dots,\allowbreak \hat{S}_{\text{DNN}_b}(t\!-\!K\!+\!1,f)]^\T$. %
The dereverberation result is computed as $Y(t,f)-\hat{\mathbf{g}}'(f)^{\H}\widetilde{\hat{\mathbf{S}}}_{\text{DNN}_b}(t,f)$, where the subtracted term is considered as the reverberation estimated by FCP.
Essentially, Eq.~(\ref{complexproj2dummy}) reverberates the target speech $\hat{S}_{\text{DNN}_b}$ estimated by the DNN using a filter per frequency to find its delayed and decayed copies.
Such copies can be reliably considered as reverberation because they are repeated signals of $\hat{S}_{\text{DNN}_b}$.
By absorbing $\hat{S}_{\text{DNN}_b}(t,f)$ into $\widetilde{\hat{\mathbf{S}}}_{\text{DNN}_b}(t,f)$, Eq.~(\ref{complexproj2dummy}) can be equivalently reformulated as
\begin{align}\label{complexproj2}
\underset{\mathbf{g}(f)}{{\text{argmin}}} \sum_{t} \frac{|Y(t,f)-\mathbf{g}(f)^{\H}\ \widetilde{\hat{\mathbf{S}}}_{\text{DNN}_b}(t,f)|^2}{\hat{\lambda}(t,f)},
\end{align}
whose goal is to filter the DNN-estimated target speech to approximate reverberant target speech $X$, assuming $V$ is weak enough.
The dereverberation result is obtained as
\begin{align}
\hat{S}_{\text{FCP}}(t,f)=Y(t,f)-\Big(\hat{\mathbf{g}}(f)^{\H}\widetilde{\hat{\mathbf{S}}}_{\text{DNN}_b}(t,f)-\hat{S}_{\text{DNN}_b}(t,f)\Big),
\label{eq:s_fcp}
\end{align}
where $\hat{\mathbf{g}}(f)^{\H}\widetilde{\hat{\mathbf{S}}}_{\text{DNN}_b}(t,f)$ is an estimate of reverberant target speech $X(t,f)$, and $\hat{\mathbf{g}}(f)^{\H}\widetilde{\hat{\mathbf{S}}}_{\text{DNN}_b}(t,f)-\hat{S}_{\text{DNN}_b}(t,f)$ the estimated reverberation of the target speaker.
Note that Eq.~\eqref{eq:s_fcp} can be rewritten as
\begin{align}
\hat{S}_{\text{FCP}}(t,f)=\hat{S}_{\text{DNN}_b}(t,f) + \Big(Y(t,f)-\hat{\mathbf{g}}(f)^{\H}\widetilde{\hat{\mathbf{S}}}_{\text{DNN}_b}(t,f)\Big),
\label{eq:s_fcp_alt}
\end{align}
which can be interpreted as adding to the initial target speech estimate $\hat{S}_{\text{DNN}_b}$ obtained by the DNN the residual component in $Y$ that cannot be explained by linear-filtering of $\hat{S}_{\text{DNN}_b}$.
\ZQHLAQ{Again}, the objective to be minimized is quadratic, and a closed-form solution exists. We will discuss the choices of $\hat{\lambda}(t,f)$ in Section \ref{lambda}.

\ZQHL{In contrast with Eq.~(\ref{dnnwpe}), Eq.~(\ref{complexproj2}) may be more effective at reducing early reflections, as it does not have a prediction delay.
In addition, it introduces a different form of linear prediction that can utilize both magnitude and phase produced by DNNs for dereverberation.} %

\subsection{Robustness of WPE, ICP, and FCP to Interferences}
\label{robustnessdescription}

Comparing Eqs.~(\ref{dnnwpe}), (\ref{complexproj1}), and (\ref{complexproj2}), we can see that the first two both do \textit{inverse} filtering, while the third does \textit{forward} filtering. This leads us to think that Eq.~(\ref{complexproj2}) may lead to better filter estimation for the target speaker when interference signals are present.
To see this, \ZQHL{we equivalently reformulate Eq.~(\ref{complexproj2}) in terms of $X$:  given that $Y=X+V$, 
\begin{align}\label{complexproj2robust}
&\underset{\mathbf{g}(f)}{{\text{argmin}}} \sum_{t} \frac{|X(t,f)+V(t,f)-\mathbf{g}(f)^{\H}\ \widetilde{\hat{\mathbf{S}}}_{\text{DNN}_b}(t,f)|^2}{\hat{\lambda}(t,f)} \nonumber \\
&= %
\underset{\mathbf{g}(f)}{{\text{argmin}}} \sum_{t} \frac{|X(t,f)-\mathbf{g}(f)^{\H}\ \widetilde{\hat{\mathbf{S}}}_{\text{DNN}_b}(t,f)|^2+|V(t,f)|^2}{\hat{\lambda}(t,f)} \nonumber \\
&= %
\underset{\mathbf{g}(f)}{{\text{argmin}}} \sum_{t} \frac{|X(t,f)-\mathbf{g}(f)^{\H}\ \widetilde{\hat{\mathbf{S}}}_{\text{DNN}_b}(t,f)|^2}{\hat{\lambda}(t,f)},
\end{align}
where the analysis assumes that $\hat{S}_{\text{DNN}_b}$ and $X$ are uncorrelated with $V$}, such that
\begin{align}
\sum_{t} \frac{V(t,f)^{\H}\Big(X(t,f)-\mathbf{g}(f)^{\H}\ \widetilde{\hat{\mathbf{S}}}_{\text{DNN}_b}(t,f)\Big)}{\hat{\lambda}(t,f)}\approx 0.
\end{align}
As we can see from Eq.~(\ref{complexproj2robust}), FCP essentially 
estimates the filter based on $\hat{S}_{\text{DNN}_b}$ and $X$, between which a linear-filter structure is indeed expected to exist.
This can produce a good filter estimate for the target speaker, even if the mixture contains noises or competing speakers.

A similar derivation for WPE's Eq.~(\ref{dnnwpe}) leads to \ZQHL{
\begin{align}\label{wperobust}
&\underset{\mathbf{g}(f)}{{\text{argmin}}} \sum_{t} \frac{{\scriptstyle |X(t,f)+V(t,f)-\mathbf{g}(f)^{\H} \big(\widetilde{\mathbf{X}}(t-\Delta,f)+\widetilde{\mathbf{V}}(t-\Delta,f)\big)|^2}}
{\hat{\lambda}(t, f)} \nonumber \\
&= %
\underset{\mathbf{g}(f)}{{\text{argmin}}}
\Big(
\sum_{t} \frac{|X(t,f)-\mathbf{g}(f)^{\H}\widetilde{\mathbf{X}}(t-\Delta,f)|^2}{\hat{\lambda}(t, f)} \nonumber \\
&\hspace{1.3cm}
+ \sum_{t} \frac{|V(t,f)-\mathbf{g}(f)^{\H}\widetilde{\mathbf{V}}(t-\Delta,f)|^2}
{\hat{\lambda}(t, f)}
\Big),
\end{align}}
where $\widetilde{\mathbf{X}}(t,f)=[X(t,f),X(t-1,f),\dots,X(t-K+1,f)]^\T$ and $\widetilde{\mathbf{V}}(t,f)=[V(t,f),V(t-1,f),\dots,V(t-K+1,f)]^\T$.
Eq.~(\ref{wperobust}) suggests that WPE aims at dereverberating the target speaker and non-target sources using a single filter.
This could be problematic when non-target sources are present, because the filter would also need to reduce the reverberation of non-target sources rather than focusing on dereverberating the target speaker.
When they are strong, the loss on non-target sources could dominate the numerator, and the resulting filter would be biased towards dereverberating higher-energy sources.
ICP's Eq.~(\ref{complexproj1}) suffers from the same issue.
In contrast, FCP's Eq.~(\ref{complexproj2}) aims at only removing the reverberation related to a target speaker.
This is particularly useful in multi-speaker separation, because each target speaker is convolved with a different RIR and it is thus best to compute a different filter to dereverberate each speaker.
This also means that our method does not aim at reducing the reverberation of non-target sources such as multi-source environmental noises, as it would require estimating each anechoic noise source, which is very difficult \cite{Kavalerov2019}.
We think this is fine because we will introduce in Section \ref{stack} a second DNN to leverage convolutive-prediction outputs for further dereverberation.
Note that noises should be relatively easier to be removed by DNNs than target-speech reverberation, because they exhibit more different patterns to target speech than target-speech reverberation does.
This is why we want the filter to focus on reducing the reverberation of the target speaker rather than that of the target speaker and non-target sources combined.
In other words, as long as FCP dereverberates the target, it should be fine if there are interferences left, because the second DNN can likely reduce interferences effectively.

\subsection{Choices for $\hat{\lambda}$ in Convolutive Prediction}\label{lambda}

In WPE, $\hat{\lambda}$ is introduced because the numerator $Y(t,f)-\mathbf{g}(f)^{\H}\ \widetilde{\mathbf{Y}}(t-\Delta,f)$ in Eq.~(\ref{wpe}) is an estimate of target speech, which is assumed to follow a complex Gaussian distribution with a time-varying PSD. \ZQHL{Apart from its origin as the variance of the target speech distribution,} $\hat{\lambda}(t,f)$ can \ZQHL{in practice also} be considered as a weighting term to balance the contributions of different T-F units, typically with diverse energy levels, for linear prediction.

Whether it is appropriate to use the estimated target-speech PSD as $\hat{\lambda}$ in convolutive prediction is not clear. In Eq.~(\ref{complexproj1}), the numerator is not an estimate of target speech.
Similarly, in Eq.~(\ref{complexproj2}), the numerator $Y(t,f)-\mathbf{g}(f)^{\H}\ \widetilde{\hat{\mathbf{S}}}_{\text{DNN}_b}(t,f)=V(t,f)+X(t,f)-\mathbf{g}(f)^{\H}\ \widetilde{\hat{\mathbf{S}}}_{\text{DNN}_b}(t,f)$ is a summation of the non-target sources $V(t,f)$ and the part of reverberant target speech $X(t,f)$ that cannot be approximated by $\mathbf{g}(f)^{\H}\widetilde{\hat{\mathbf{S}}}_{\text{DNN}_b}(t,f)$.
For both numerators, we could assume a Gaussian distribution, as sums of Gaussian variables are also Gaussian for independent variables, and use iterative optimization to find \ZQHL{the PSD and the filter} in the same way as in the vanilla WPE algorithm \cite{Nakatani2010}.
However, this would introduce time- and computation-consuming iterations.
We could avoid the iterations by estimating the PSD using another DNN, or more economically as another output of $\text{DNN}_1$.
Both of these approaches however increase the complexity of our system.

\ZQHL{Rather than focusing on such a probabilistic interpretation, we introduce $\hat{\lambda}$ as a weighting term that can balance T-F units with different energy for linear prediction.
Without it, the filter could be biased towards only producing good estimates for higher-energy T-F units.
One potential choice for such a $\hat{\lambda}$ is to set $\hat{Z}$ in Eq.~(\ref{wpelambda}) to $|\hat{S}_{\text{DNN}_b}|$, leading to}
\begin{align}\label{complexproj1lambda}
\hat{\lambda}(t,f)=\text{max}(\varepsilon \text{max}(|\hat{S}_{\text{DNN}_b}|^2),|\hat{S}_{\text{DNN}_b}(t,f)|^2).
\end{align}
\ZQHL{Another option is to simply use the mixture power, leading to}
\begin{align}\label{complexproj2lambda}
\hat{\lambda}(t,f)=\text{max}(\varepsilon \text{max}(|Y|^2),|Y(t,f)|^2).
\end{align}
\ZQHL{Although such settings for $\hat{\lambda}$ do not follow a probabilistic assumption, we found that they work well in our experiments. In particular, the best results were obtained with Eq.~(\ref{complexproj1lambda}) for ICP and with Eq.~(\ref{complexproj2lambda}) for FCP.
}

\subsection{Combining Convolutive-Prediction Outputs with DNN}\label{stack}

The dereverberation result produced by convolutive prediction explicitly exploits the linear-filter structure in reverberation.
Such a linear structure is assumed time-invariant for non-moving sources, and could be estimated reasonably well by using all the frames in the mixture, as long as the mixture is reasonably long.
It could leverage information complementary to, and hence improve upon, plain DNN based dereverberation, where the DNN output at each T-F unit might not strictly respect the linear-filter structure.
Furthermore, because the convolutive-prediction results are obtained based on a time-invariant filter, it is likely that they are not as good as DNN outputs in terms of speech enhancement metrics.

We hence consider using the convolutive-prediction outputs as extra features to train another DNN for better dereverberation.
The input feature is a concatenation of the RI components of $Y$, the estimated target speech produced by the first DNN $\hat{S}_{\text{DNN}_1}$, and a linear prediction result such as $\hat{S}_{\text{WPE}}$, $\hat{S}_{\text{ICP}}$, or $\hat{S}_{\text{FCP}}$.
We will compare the performance of using different convolutive-prediction features for DNN training. This second network's output is denoted as $\hat{S}_{\text{DNN}_2}$.

As a baseline, we consider only using $Y$ and $\hat{S}_{\text{DNN}_1}$ as features to train the second DNN, i.e., not incorporating linear-prediction results.
This stacking approach was conventionally perceived as a model ensembling technique, often used in early deep learning based speech enhancement studies that operate in the magnitude domain \cite{Narayanan2014, Nie2014, Zhang2016, Wang2017}.
Here, we point out that it is a very valid idea closely related to the proposed technique.
Our insight is that since our second DNN takes in the RI components of $Y$ and $\hat{S}_{\text{DNN}_1}$ (not just their magnitudes) as inputs to predict $S$, this stacking approach could implicitly identify copies of $\hat{S}_{\text{DNN}_1}$ contained in $Y$ based on supervised learning, rather than by using $Y$ and $\hat{S}_{\text{DNN}_1}$ for explicit linear regression.
In other words, the DNN could learn to exploit the linear and non-linear information in $Y$ and $\hat{S}_{\text{DNN}_1}$ to best predict target speech in a data-driven way.
Although this two-DNN stacking system produces clear improvement over the one-DNN system, we will show in our experiments that incorporating convolutive-prediction results to train the second DNN leads to further improvement.

\subsection{Run-Time Iterative Processing}\label{iteration}
As $\hat{S}_{\text{DNN}_2}$ is likely better than $\hat{S}_{\text{DNN}_1}$ and thus potentially leads to better filter estimation, we can use it to do another pass of convolutive prediction. We then combine $\hat{S}_{\text{DNN}_2}$, $Y$, and the new projection results as input features for the second DNN to estimate target speech again. Note that, in this work, we do not train the second DNN based on this second pass output, but that could be considered in future work. %

We point out that the big picture of our approach is essentially an iterative strategy for blind deconvolution, which is a non-convex problem in nature and is difficult to solve without assuming any knowledge of the source signal or the filter \ZQHLAQ{\cite{Levin2011, blinddeconvolution}}.
Our approach first uses a DNN, which is known to be good at modeling speech patterns \cite{WDL2018}, to estimate the direct-path signal.
Given an estimate of the direct-path signal, estimating the RIR becomes an easier convex, linear regression problem.
We here estimate the RIR via FCP in the T-F domain.
Given an estimate of the RIR, estimating the direct-path signal also becomes easier.
In this work, we use the estimated RIR to derive extra features for another DNN to better predict the direct-path signal.
We can iterate this process to gradually improve the estimation of the RIR and the direct-path signal.

\section{DNN Configurations}\label{dnndescription}
Our DNNs operate in the complex T-F domain via complex spectral mapping. %
This section provides the detailed DNN setup for speech dereverberation and speaker separation, as well as DNN architectures.

\subsection{Complex Spectral Mapping}

Our DNNs predict the RI components of the direct-path signal from the mixture RI components.
This approach and the related complex ratio masking technique \cite{Williamson2016} have shown strong performance in tasks such as speech dereverberation \cite{Wang2020b}, speech enhancement \cite{Fu2017, Wang2020a} and speaker separation \cite{Liu2019,Wang2020c}.

\subsection{Speech Dereverberation}
For speech dereverberation, following \cite{Wang2020b} we define the loss function on the predicted RI components %
\begin{align}\label{enhloss}
\mathcal{L}^{(b)}_{\text{Enh, RI}} &= \| \hat{R}^{(b)} - \text{Real}(S)\|_1 + \| \hat{I}^{(b)} - \text{Imag}(S)\|_1,
\end{align}
where $\hat{R}^{(b)}$ and $\hat{I}^{(b)}$ are the predicted RI components produced by using linear activations in the output layer, $b\in\{1,2\}$ denotes which DNN produces the estimates (as we have two DNNs in our best performing system), $\text{Real}(\cdot)$ and $\text{Imag}(\cdot)$ extract RI components, and $\| \cdot\|_1$ computes the $L_1$ norm. Following \cite{Wang2020b, Wang2020a}, we further add a loss on the resulting magnitude, leading to 
\begin{align}\label{enhlossri+mag}
\mathcal{L}^{(b)}_{\text{Enh, RI+Mag}} &= 
\mathcal{L}^{(b)}_{\text{Enh, RI}} + \Big\| \sqrt{\hat{R}^{(b)^2}+\hat{I}^{(b)^2}} - |S|\Big\|_1.
\end{align}
The enhancement result is obtained as $\hat{S}_{\text{DNN}_b}=\hat{R}^{(b)}+j\hat{I}^{(b)}$, where $j$ denotes the imaginary unit.
Inverse STFT is then applied to get the estimated time-domain signal.

We point out that the trained DNN essentially does complex-domain inverse filtering, similarly to WPE and ICP, but here we use supervised learning to learn non-linear inverse filters based on a large receptive field, which is possible thanks to the use of DNNs.
\ZQHLAQ{Note that complex-domain approaches \cite{Wang2020b} typically achieve better dereverberation than magnitude-domain approaches \cite{Han2015, Ernst2018, Zhao2020}. }

\subsection{Speaker Separation}
For speaker separation, we also define the loss based on the predicted RI components, but we additionally use utterance-wise PIT \cite{R.Hershey2016, Isik2016, Kolbak2017} to address the label-permutation problem. Here, we introduce a speaker index $c\in \{1,\dots,C\}$ to differentiate between the $C$ speakers $S(1),\dots,S(C)$. The loss function is defined as
\begin{align}\label{pitloss}
\mathcal{L}^{(1)}_{\text{PIT}} = & \min_{\pi \in \mathcal{P}} \sum_{c} \bigs(
\| \hat{R}^{(1)}(\pi(c)) - \text{Real}(S(c))\|_1 \nonumber \\
&\hspace{1.5cm}+ \| \hat{I}^{(1)}(\pi(c)) - \text{Imag}(S(c))\|_1
\bigs),
\end{align}
where $\mathcal{P}$ is the set of permutations on $\{1,\dots,C\}$. 
The separation result is obtained as $\hat{S}_{\text{DNN}_1}(c)=\hat{R}^{(1)}(c)+j\hat{I}^{(1)}(c)$.
We find that adding to $\mathcal{L}^{(1)}_{\text{PIT}}$ a loss on the sum of the target speech estimates improves separation especially in noisy-reverberant conditions. That loss is defined as
\begin{align}\label{sumpitloss}
\mathcal{L}^{(1)}_{\text{sumPIT}} =& \bigs\| \sum_{c} \hat{R}^{(1)}(c) - \text{Real}\bigs(\sum_{c}S(c)\bigs)\bigs\|_1 \nonumber \\
&+ \bigs\| \sum_{c} \hat{I}^{(1)}(c) - \text{Imag}\bigs(\sum_{c}S(c)\bigs)\bigs\|_1.
\end{align}
We train $\text{DNN}_1$ using either $\mathcal{L}^{(1)}_{\text{PIT+sumPIT}}=\mathcal{L}^{(1)}_{\text{PIT}}+\mathcal{L}^{(1)}_{\text{sumPIT}}$ or $\mathcal{L}^{(1)}_{\text{PIT}}$.
Note that we use superscript $(1)$ here, as PIT is only used for $\text{DNN}_1$.

\begin{figure}
  \centering  
  \includegraphics[width=9cm]{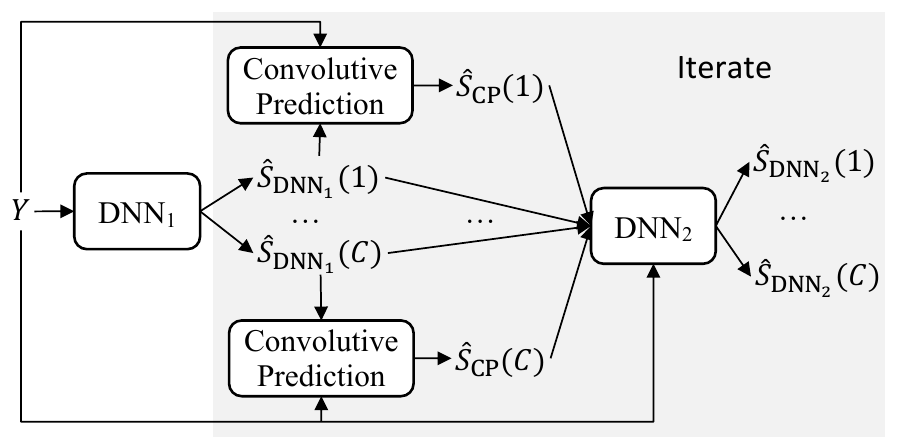} \vspace{-0.6cm} \\
  \caption{\footnotesize{Illustration of the ``all speakers'' setup of the speaker separation system, where $\text{DNN}_2$ enhances all speakers simultaneously.}}
  \label{systemspksep1} \vspace{-0.4cm}
\end{figure}

\begin{figure}
  \centering
  \includegraphics[width=9cm]{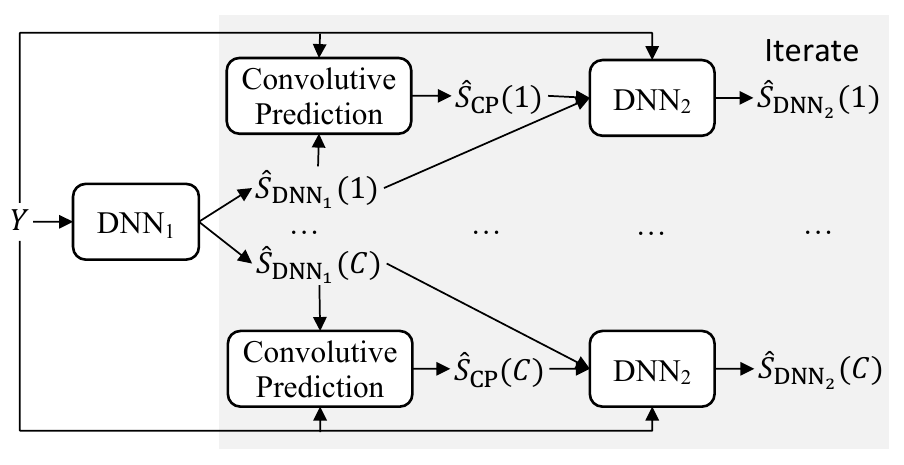} \vspace{-0.6cm} \\  
  \caption{\footnotesize{Illustration of the ``per speaker'' setup of the speaker separation system, where $\text{DNN}_2$ enhances each speaker independently.}}
  \label{systemspksep2} \vspace{-0.4cm}
\end{figure}

For $\text{DNN}_2$, we can just use an enhancement network to enhance all the speakers, as $\text{DNN}_1$ has resolved the permutation problem.
The loss function on each speaker follows Eqs.~(\ref{enhloss}) or (\ref{enhlossri+mag}).
We think that training the second network in an enhancement way should produce better performance than training it with PIT, as the network already knows the right permutation.
We consider two ways to train the enhancement network.
In the ``all speakers'' setup, shown in Fig.~\ref{systemspksep1}, we predict all $C$ target speakers simultaneously by using a concatenation such as $\big[Y, \hat{S}_{\text{DNN}_1}(1),\dots,\hat{S}_{\text{DNN}_1}(C),\hat{S}_{\text{FCP}}(1),\dots,\hat{S}_{\text{FCP}}(C)\big]$ as input to predict $\big[S(1),\dots,S(C)\big]$.
In our experiments, we use the default speaker order in each mixture and do not shuffle the speakers when training the second DNN.
In the ``per speaker'' setup, illustrated in Fig.~\ref{systemspksep2}, we predict target speakers one by one by using for example $\big[Y, \hat{S}_{\text{DNN}_1}(c),\hat{S}_{\text{FCP}}(c)\big]$ as input to predict $S(c)$. The downside is that the DNN needs to run $C$ times at run time, once for each speaker. 
In our experiments, the ``per speaker'' setup produces clearly better results. \ZQHLAQ{We think that there are three possible explanations: (1) each speaker is convolved with a different RIR, so it is best to do inverse filtering separately for each speaker, similarly to the proposed convolutive prediction; (2) DNNs are better at modeling the pattern of a single target speaker than that of multiple target speakers combined; and (3) there is more training data for the second network in the ``per speaker'' case.}

\begin{figure}
  \centering  
  \includegraphics[width=8cm]{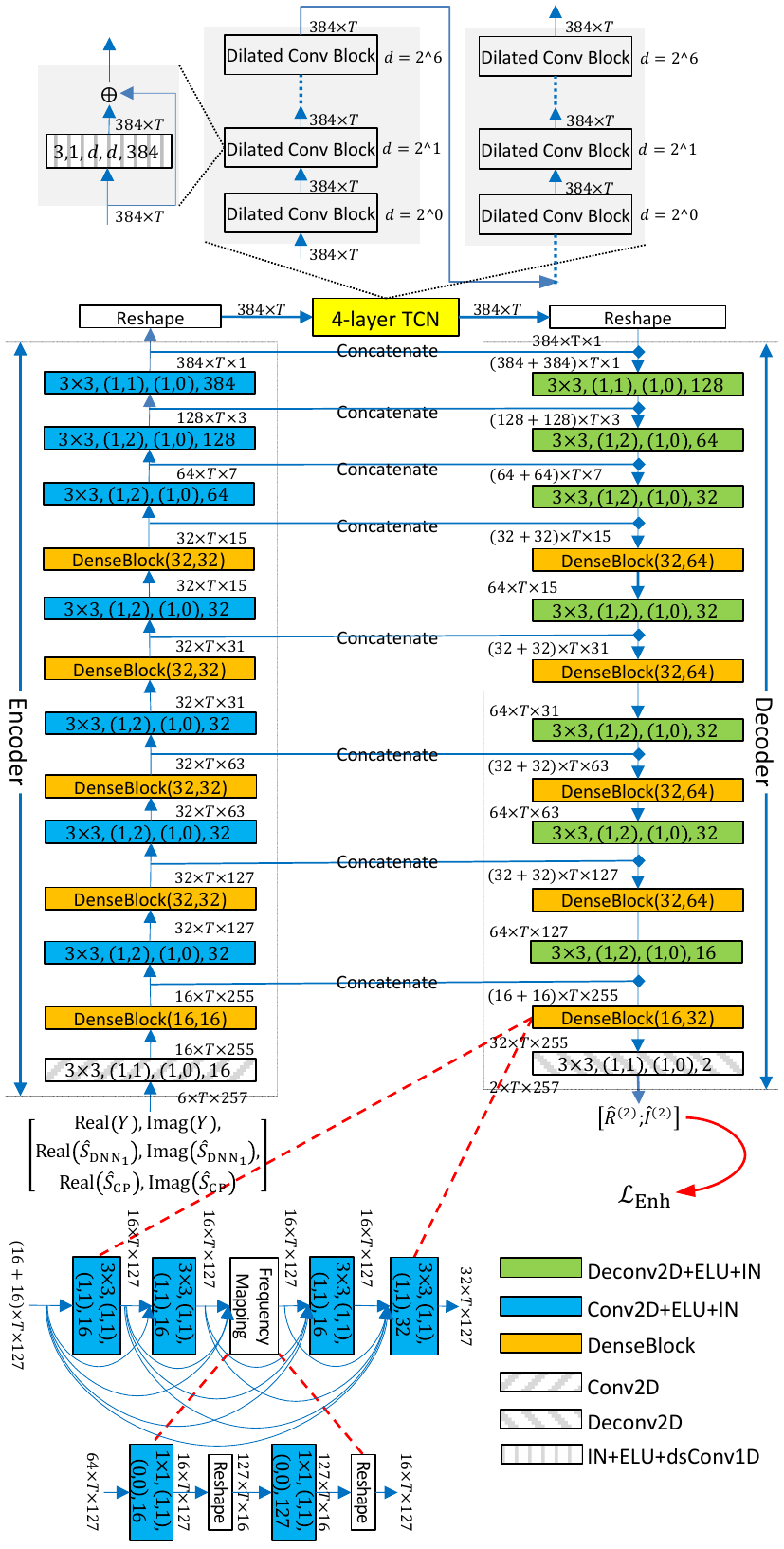} \vspace{-0.3cm} \\  
  \caption{\footnotesize{Example network architecture of $\text{DNN}_2$ for dereverberation.
The tensor shape after each encoder-decoder block is in the format:
\textit{featureMaps×timeSteps×frequencyChannels}. Each one of Conv2D, Deconv2D, Conv2D+ELU+IN, and Deconv2D+ELU+IN blocks is specified in the format: \textit{kernelSizeTime×kernelSizeFreq, (stridesTime, stridesFreq), (paddingsTime, paddingsFreq), featureMaps}.
Each DenseBlock($g_1$,$g_2$) contains five Conv2D+ELU+IN blocks
with growth rate $g_1$ for the first four layers and $g_2$ for the last layer.
Following \cite{Liu2019}, we include a frequency mapping layer in the middle of each denseblock.
The tensor shape after each TCN block is in the format: \textit{featureMaps×timeSteps}.
Each IN+ELU+dsConv1D block is specified in
the format: \textit{kernelSizeTime, stridesTime, paddingsTime, dilationTime,
featureMaps}.}}
  \label{dnnfigure} \vspace{-0.4cm}
\end{figure}

\subsection{Network Architecture}

The network architecture of $\text{DNN}_2$ is shown in Fig.~\ref{dnnfigure}. 
$\text{DNN}_1$ also uses the same architecture, but only uses the RI components of the mixture as input.
It is a temporal convolutional network (TCN) \cite{Bai2018} sandwiched by a U-Net \cite{Ronneberger2015}. We insert DenseNet blocks \cite{Huang2017} at multiple frequency scales in the encoder and decoder. The motivation of this  network  design is that U-Net can maintain local fine-grained structure via its skip connections and model contextual information along frequency through down- and up-sampling, TCN can leverage long-range  information by using dilated convolutions along time, and DenseNet blocks encourage feature reuse and improve discriminability.
More specifically, the encoder contains one two-dimensional (2D) convolution, and seven convolutional blocks, each with 2D convolution, exponential linear units (ELU) non-linearity, and instance normalization (IN), for down-sampling.
The decoder includes seven blocks of 2D deconvolution, ELU, and IN, and one 2D deconvolution, for up-sampling.
The TCN contains four layers, each of which has seven dilated convolutional blocks.
We use one one-dimensional depth-wise separable convolution, denoted as dsConv1D, in each dilated convolutional block to reduce the number of parameters.

We stack the RI components of different input signals as features maps in the network input and output. Linear activations are used in the output layer.

\section{Experimental Setup}\label{setup}

We evaluate the proposed algorithms on three tasks: speech dereverberation with weak stationary noise, two-speaker separation in reverberant conditions with white noise, and two-speaker separation in reverberant conditions with challenging non-stationary noise.
These three tasks are progressively more difficult, as we consider competing speakers and challenging noises that are known to be detrimental to linear prediction. 
This section describes the dataset used for each task, the hyper-parameter settings, the evaluation metrics, and the baseline systems.

\subsection{Dataset for Dereverberation}

For speech dereverberation, we train our models on a simulated reverberant dataset with weak air-conditioning noise.
In addition to evaluating the trained models on our simulated test set, we apply them directly to the REVERB corpus \cite{Kinoshita2016} to show their effectiveness in dealing with real-recorded noisy-reverberant utterances.

The clean signals for simulation are from the WSJCAM0 corpus.
It contains 7861, 742, and 1088 utterances in its training, validation, and test set, respectively.
We use them to simulate 39305 (7861$\times$5), 2968 (742$\times$4), and 3264 (1088$\times$3) noisy-reverberant mixtures as our training, validation, and test sets, respectively.
The data spatialization process follows \cite{Wang2020d}, where, for each utterance, we randomly sample a room with random room characteristics and speaker and microphone locations, using the RIR generator proposed in \cite{Scheibler2018}.
The speaker-to-microphone distance is sampled from the range $[0.75, 2.5]$ m.
The reverberation time (T60) is drawn from the range $[0.2, 1.3]$ s.
For each utterance, a diffuse air-conditioning noise is sampled from the REVERB corpus \cite{Kinoshita2016} and added to the reverberant speech.
The signal-to-noise ratio between the anechoic speech and the noise is sampled from the range $[5, 25]$ dB.
The sampling rate is 16 kHz.
We denote this simulated dataset as ``Dereverb Data I''.

To show the generalizability of our trained models to realistic reverberant recordings, we apply them, without retraining, to the ASR tasks of REVERB.
The test mixtures are from real recordings made in rooms with T60 around 0.7 s and with speaker-to-microphone distance around 1 m in the near-field case and 2.5 m in the far-field case. 
The recorded noise is diffuse air-conditioning noise and is weak.

We use the official REVERB corpus \cite{Kinoshita2016} and the most recent Kaldi recipe \cite{Povey2011} to build our ASR backend.
It is trained using the noisy-reverberant speech plus clean source signals of REVERB. We follow a plug-and-play approach for ASR, where enhanced time-domain signals are directly fed into the backend for decoding.

\subsection{Dataset for Reverberant Speaker Separation}

We utilize the six-channel SMS-WSJ dataset \cite{Drude2019}, which contains simulated two-speaker mixtures in reverberant conditions.
The clean speech is sampled from the WSJ0 and WSJ1 datasets.
The corpus contains 33561, 982, and 1332 two-speaker mixtures for training, validation, and testing, respectively.
The speaker-to-array distance is sampled from  the  range $[1.0, 2.0]$ m, and the T60 is drawn from the range $[0.2, 0.5]$ s.
A weak white noise is added to simulate microphone noises.
The energy level between the sum of the reverberant target speech signals and the noise is sampled from the range $[20, 30]$ dB.
The sampling rate is 8 kHz.
We only use the first channel for training and evaluation.
We use the direct sound as the training target and perform both dereverberation and separation. 
Our study aims at removing all the reflections. This is different from the default setup in SMS-WSJ, which does not aim at reducing early reflections.

For ASR, we use the default Kaldi-based backend acoustic model provided in SMS-WSJ \cite{Drude2019}, trained using single-speaker noisy-reverberant speech as inputs and the state alignment of its corresponding direct-path signal as labels.
The signals at the first, third, and fifth channels are used for training the acoustic model.
A task-standard trigram language model is used for decoding.

\subsection{Dataset for Noisy-Reverberant Speaker Separation}

We conduct our experiments on the noisy-reverberant WHAMR!~dataset \cite{Maciejewski2020}, originally designed for noisy-reverberant binaural two-speaker separation.
It re-uses the clean two-speaker mixtures in the wsj0-2mix dataset \cite{R.Hershey2016}, but reverberates each clean signal and adds non-stationary environmental noise recorded in WHAM!~\cite{Wichern2019}.
The T60 is randomly sampled from the range $[0.2, 1.0]$ s.
The signal-to-noise ratio between the louder speaker and noise is drawn from the range $[-6, 3]$ dB.
The energy level between the two speakers in each mixture is sampled from the range $[-5, 5]$ dB.
The speaker-to-array distance is sampled from the range $[0.66, 2.0]$ m.
There are 20000, 5000, and 3000 binaural mixtures in the training, validation, and test set, respectively.
We use the \textit{min} and 8 kHz version of the corpus.
We only use the first channel for training and evaluation.
We aim at joint dereverberation, denoising, and speaker separation. The direct-path signal of each speaker is used as the reference for metric computation.

\subsection{Miscellaneous Configurations}

For STFT, the window length is 32 ms, hop size is 8 ms, and the analysis window is the square root of the Hann window. For 16 kHz sampling rate, a 512-point fast Fourier transform (FFT) is applied to extract 257-dimensional STFT features, while a 256-point FFT is used to extract 129-dimensional features for 8 kHz sampling rate. 
No sentence- or global-level mean-variance normalization is performed on input features. 
For each mixture, we normalize its sample variance to one before any processing.
Note that during training, the target signal needs to be scaled by the same factor used for scaling the mixture.

For WPE and DNN-WPE, the number of filter taps $K$ is set to 37 and the prediction delay $\Delta$ is set to 3, following \cite{Kinoshita2016, Kinoshita2017}.
The iteration number in the vanilla WPE is set to 3.
No PSD context \cite{Drude2020} is used.
Note that, based on the validation set, we compared setting $K$ and $\Delta$ to 40 and 0, 39 and 1, 38 and 2, 37 and 3, and 36 and 4, and found that setting them to 37 and 3 works best across our datasets.
For convolutive prediction, no prediction delay is used and $K$ is set to 40, leading to the same amount of context as in WPE.
This amounts to 344 ($=(40-1)\times 8+32$) ms filter length in the time domain.
We increased $K$ to up to 125, which corresponds to up to $1.0$ s RIR length.
This leads to an increase in the amount of computation spent in the linear regression step, but we did not observe significant differences in the evaluation scores.
This is possibly because the RIRs used in this study have their energy mostly in the $0.35$ s range after the peak impulse. The floor value
$\varepsilon$ in Eqs.~(\ref{wpelambda}), (\ref{complexproj1lambda}), and (\ref{complexproj2lambda}) is set to either $1.0$, meaning that no weighting is used, or $0.001$, meaning that the PSD at each T-F unit should be at most $-30$ dB lower than the T-F unit with the highest energy.

\subsection{Evaluation Metrics}

For all the tasks, our major evaluation metrics include the scale-invariant signal-to-distortion ratio (SI-SDR) \cite{LeRoux2019}, which measures the quality of time-domain sample-level predictions, extended short-time objective intelligibility (eSTOI) \cite{H.Taal2011}, and perceptual evaluation of speech quality (PESQ) scores.
For PESQ, we report narrow-band MOS-LQO scores based on the ITU P.862.1 standard \cite{P862.1} using the \textit{python-pesq} toolkit\footnote{https://github.com/ludlows/python-pesq, v0.0.2.}.
We report word error rates (WER) for ASR.
\ZQHLAQ{In all the tasks, the direct-path signal, which physically represents the target signal captured by a far-field microphone in anechoic conditions, is always used as the reference for metric computation.}

\ZQHLAQ{
In addition to SI-SDR, we also compute BSS-Eval SDR \cite{Vincent2006, Vincent2006code}.
We emphasize that SDR manipulates the reference signal using a time-invariant 512-tap filter, which is 32 ms long for 16 kHz sampling rate and 64 ms for 8 kHz, to best approximate the estimated signal before computing an SNR-like score, while SI-SDR only using a one-tap filter.
As a result, SDR is limited in its ability to measure whether early reflections are removed, as the 512-tap filter would re-create a \textit{reverberant} signal for metric computation.
For systems that are not able to, or not designed to, reduce early reflections, SDR can still produce a relatively high score, but our goal in this paper is to remove any reflections, and therefore the interpretation of the SDR scores in this paper is very tricky and requires caution.
On the other hand, SI-SDR \cite{LeRoux2019} only modifies the reference by a scaling factor (or a one-tap time-invariant filter) before computing an SNR-like score: it tolerates a gain difference between the separated signal and the reference, but significantly penalizes phase errors.
While this may be unfair to algorithms that do not attempt to or are not able to preserve time alignment, this is not an issue here as the methods considered in this study do output a time-aligned estimate of the direct-path signal (see our discussion in the last paragraph of Section~\ref{reducingearlyresults}).
We emphasize that preserving time alignment is desired in many application scenarios such as active noise control, %
spatial cue preservation, some beamforming algorithms, and real-time speech enhancement.}

Note that, in this study, we obtain the direct-path RIR by setting the T60 parameter of the RIR generator to zero, and then convolve it with the dry signal to obtain the direct-path signal.
In the literature, some studies only consider the peak impulse of the entire RIR (i.e., a single pulse) as the direct-path RIR \cite{Delfarah2020, Zhao2020}.
As is pointed out in \cite{Wang2020b}, this peak impulse approach has some issues.
Because we rely on discrete-time digital signal processing, the direct-path impulse likely does not happen exactly at a sample instance.
This is particularly problematic in the context of microphone array processing.
If two microphones are placed very closely in space, this peak-impulse approach may lead to the same direct-path RIR for the two microphones.
The resulting direct-path signals at the two microphones would be exactly the same, not exhibiting any phase difference.
Even if the two direct-path RIRs differ by, say, one sample shift in time, the phase difference between the resulting direct-path signals would be discretized and could lead to problems for later spatial processing.

\begin{table*}[t]
\footnotesize
\centering
  \sisetup{table-format=2.2,round-mode=places,round-precision=2,table-number-alignment = center,detect-weight=true,detect-inline-weight=math}
\caption{\footnotesize{SI-SDR (dB), SDR (dB), and PESQ results on test set of \ZQHL{Dereverb Data I}, and WER (\%) on \ZQHLAQ{Real Data of} REVERB.}}
\label{resultsdereverb}
\setlength{\tabcolsep}{2.5pt}
\begin{tabular}{lcccc!{\enspace}S[table-format=2.1,round-precision=1]!{\enspace}S[table-format=2.1,round-precision=1]!{\enspace}S[table-format=1.2]!{\enspace}SSS!{\enspace}SSS}
\toprule %
& \multirow{2}{*}[-6.5pt]{\begin{tabular}[c]{@{}c@{}}$\text{DNN}_1/\text{DNN}_2$ \\ predicts?\end{tabular}} & \multirow{2}{*}[-6.5pt]{\begin{tabular}[c]{@{}c@{}}$\text{DNN}_2$ \\ loss\end{tabular}} & & & & & & \multicolumn{3}{c}{WER on val. set} & \multicolumn{3}{c}{WER on test set} \\
\cmidrule(lr{9pt}){9-11} \cmidrule(lr){12-14} %
\multicolumn{1}{l}{Approaches} &  & & $\hat{\lambda}$ & $\varepsilon$ & \hspace{-.35em} {SI-SDR} & \hspace{-.35em} {SDR} & \hspace{-.25em}{PESQ} & \multicolumn{1}{c}{Near} & \multicolumn{1}{c}{Far} & {Avg.} & \multicolumn{1}{c}{Near} & \multicolumn{1}{c}{Far} & {Avg.} \\
\midrule %
Unprocessed & - & - & - & - & -3.6 & 2.6729 & 1.64 & 15.35 & 16.88 & 16.11 & 17.09 & 17.29 & 17.19 \\ %
\midrule
vWPE (1iter) & - & - & (\ref{complexproj2lambda}) & $0.001$ & -1.6 & 5.30566 & 1.74 & 16.59 & 17.63 & 17.11 & 13.67 & 16.34 & 15.00 \\ %
vWPE (3iter) & - & - & (\ref{complexproj2lambda}) & $0.001$ & -1.4 & 5.738 & 1.74 & 16.66 & 17.57 & 17.12 & 14.02 & 16.88 & 15.45 \\

\midrule %

$\text{DNN}_{1}$ & d+e/- & & - & - & -1.6 & 7.06669 & 2.13 & 15.41 & 18.39 & 16.90 & 17.37 & 16.31 & 16.84 \\ %
\quad $\text{DNN}_{1}$+$\text{DNN}_{2}$ & d+e/d & RI & (\ref{wpelambda}) & - & 8.8 & 10.2235 & 2.77 & 12.48 & 14.29 & 13.38 & 11.31 & 11.61 & 11.46 \\ %
\quad $\text{DNN}_{1}$+WPE & d+e/- & - & (\ref{wpelambda}) & $0.001$ & -1.5 & 6.300558 & 1.75 & 15.78 & 17.70 & 16.74 & 14.31 & 16.04 & 15.18 \\ %
\quad $\text{DNN}_{1}$+WPE+$\text{DNN}_{2}$ & d+e/d & RI & (\ref{wpelambda}) & $0.001$ & 9.6 & 11.3550 & 2.95 & 11.67 & 12.78 & 12.22 & 10.19 & 9.76 & 9.97 \\

\midrule %

$\text{DNN}_{1}$ & d+e+v/- & - & - & - & -1.8 & 6.2966 & 1.95 & 14.78 & 17.16 & 15.97 & 18.27 & 17.15 & 17.71 \\ %
\quad $\text{DNN}_{1}$+$\text{DNN}_{2}$ & d+e+v/d & RI & (\ref{wpelambda}) & - & 8.2 & 9.78436 & 2.69 & 12.91 & 13.53 & 13.22 & 11.56 & 11.99 & 11.78 \\ %
\quad $\text{DNN}_{1}$+WPE & d+e+v/- & - & (\ref{wpelambda}) & $0.001$ & -1.5 & 6.24569 & 1.75 & 15.53 & 17.98 & 16.75 & 13.80 & 16.04 & 14.92 \\ %
\quad $\text{DNN}_{1}$+WPE+$\text{DNN}_{2}$ & d+e+v/d & RI & (\ref{wpelambda}) & $0.001$ & 9.2 & 10.98198 & 2.93 & 11.67 & 12.44 & 12.05 & 9.13 & 10.03 & 9.58 \\

\midrule %

$\text{DNN}_{1}$ & d/- & - & - & - & 8.2 & 9.764 & 2.65 & 12.48 & 14.56 & 13.52 & 11.69 & 11.17 & 11.43 \\ %
\quad $\text{DNN}_{1}$+$\text{DNN}_{2}$ & d/d & RI & - & - & 9.1 & 10.666 & 2.82 & 11.85 & 12.58 & 12.21 & 10.80 & 10.84 & 10.82 \\ %
\quad $\text{DNN}_{1}$+WPE & d/- & - & (\ref{complexproj1lambda}) & $0.001$ & -1.0 & 6.375 & 1.74 & 15.41 & 17.91 & 16.66 & 14.21 & 15.83 & 15.02 \\ %
\quad $\text{DNN}_{1}$+WPE+$\text{DNN}_{2}$ & d/d & RI & (\ref{complexproj1lambda}) & $0.001$ & 11.2 & 13.1100 & 3.12 & 12.29 & 12.58 & 12.43 & 9.42 & 9.52 & 9.47 \\ %
\quad $\text{DNN}_{1}$+(WPE+$\text{DNN}_{2}$)$\times$2 & d/d & RI & (\ref{complexproj1lambda}) & $0.001$ & 11.3 & 13.2798 & 3.18 & 11.17 & 12.44 & 11.80 & 9.55 & 9.59 & 9.57 \\ 

\midrule %

\quad $\text{DNN}_{1}$+vWPE (1iter)+$\text{DNN}_{2}$ & d/d & RI & (\ref{complexproj2lambda}) & $0.001$ & 10.856 & 12.621 & 3.113 & 11.04 & 12.30 & 11.67 & 9.33 & 10.23 & 9.78 \\ %

\midrule %

\quad $\text{DNN}_{1}$+ICP & d/- & - & (\ref{complexproj1lambda}) & $1.0$ & 3.2 & 5.927373 & 1.78 & 20.02 & 22.35 & 21.19 & 16.64 & 20.46 & 18.55 \\ %
\quad $\text{DNN}_{1}$+ICP+$\text{DNN}_2$ & d/d & RI & (\ref{complexproj1lambda}) & $1.0$ & 11.3 & 13.3936 & 3.10 & 10.85 & 13.19 & 12.02 & 8.85 & 9.86 & 9.36 \\ %
\quad $\text{DNN}_{1}$+(ICP+$\text{DNN}_2$)$\times$2 & d/d & RI & (\ref{complexproj1lambda}) & $1.0$ & 11.3 & 13.45342 & 3.11 & 10.92 & 13.67 & 12.29 & 9.36 & 10.03 & 9.70 \\ %

\quad $\text{DNN}_{1}$+ICP & d/- & - & (\ref{complexproj1lambda}) & $0.001$ & 0.7 & 5.6672 & 1.77 & 17.53 & 20.71 & 19.12 & 15.01 & 18.60 & 16.80 \\ %
\quad $\text{DNN}_{1}$+ICP+$\text{DNN}_2$ & d/d & RI & (\ref{complexproj1lambda}) & $0.001$ & 10.7 & 12.6725 & 3.03 & 11.17 & 11.83 & 11.50 & 8.46 & 10.40 & 9.43 \\ %

\quad $\text{DNN}_{1}$+ICP & d/- & - & (\ref{complexproj2lambda}) & $0.001$ & 1.9 & 5.1685 & 1.75 & 17.22 & 20.98 & 19.10 & 14.53 & 18.26 & 16.39 \\ %
\quad $\text{DNN}_{1}$+ICP+$\text{DNN}_2$ & d/d & RI & (\ref{complexproj2lambda}) & $0.001$ & 11.07 & 13.0926 & 3.070 & 11.10 & 12.99 & 12.04 & 9.39 & 10.03 & 9.71 \\

\midrule %

\quad $\text{DNN}_{1}$+FCP & d/- & - & (\ref{complexproj1lambda}) & $0.001$ & 2.8 & 3.8504 & 1.80 & 20.27 & 21.46 & 20.87 & 18.81 & 17.22 & 18.02 \\ %
\quad $\text{DNN}_{1}$+FCP+$\text{DNN}_2$ & d/d & RI & (\ref{complexproj1lambda}) & $0.001$ & 11.946 & 14.0933 & 3.165 & 10.61 & 11.28 & 10.95 & 8.88 & 9.22 & 9.05 \\ %

\quad $\text{DNN}_{1}$+FCP & d/- & - & (\ref{complexproj2lambda}) & $1.0$ & 3.6 & 4.45548 & 1.82 & 18.22 & 21.19 & 19.70 & 18.14 & 16.95 & 17.55 \\ %
\quad $\text{DNN}_{1}$+FCP+$\text{DNN}_2$ & d/d & RI & (\ref{complexproj2lambda}) & $1.0$ & 11.9 & 14.0956 & 3.15 & 9.86 & 12.71 & 11.29 & 8.91 & 9.62 & 9.27 \\ %

\quad $\text{DNN}_{1}$+FCP & d/- & - & (\ref{complexproj2lambda}) & $0.001$ & 3.0 & 4.26848 & 1.82 & 17.34 & 20.23 & 18.79 & 16.74 & 16.61 & 16.67 \\ %
\quad $\text{DNN}_{1}$+FCP+$\text{DNN}_2$ & d/d & RI & (\ref{complexproj2lambda}) & $0.001$ & 12.3 & 14.498 & 3.18 & 9.73 & 11.83 & 10.78 & 8.40 & 8.95 & 8.68 \\ %
\quad $\text{DNN}_{1}$+(FCP+$\text{DNN}_2$)$\times$2 & d/d & RI & (\ref{complexproj2lambda}) & $0.001$ & \bfseries 13.3 & \bfseries 15.83651 & 3.30 & 9.11 & 12.03 & 10.57 & 8.21 & 8.74 & 8.48 \\

\midrule %

\quad $\text{DNN}_{1}$+FCP+$\text{DNN}_2$ & d/d & RI+Mag & (\ref{complexproj2lambda}) & $0.001$ & 11.8 & 13.84355 & 3.39 & 8.67 & \bfseries 9.77 & \bfseries 9.22 & 7.82 & \bfseries 8.00 & 7.91 \\ %
\quad $\text{DNN}_{1}$+(FCP+$\text{DNN}_2$)$\times$2 & d/d & RI+Mag & (\ref{complexproj2lambda}) & $0.001$ & 12.7 & 14.9900 & \bfseries 3.46 & \bfseries 8.36 & 10.39 & 9.38 & \bfseries 7.63 & 8.17 & \bfseries 7.90 \\

\midrule\midrule

8ch WPE+BeamformIt! (in Kaldi) & - & - & - & - & {-} & {-} & {-} & 10.85 & 9.36 & 10.11 & 8.85 & 8.74 & 8.79 \\

\bottomrule %
\end{tabular}%
\vspace{-0.4cm}
\end{table*}

\subsection{Benchmark Systems}

For dereverberation, we compare the proposed algorithms with WPE, DNN-WPE, and their variants, either by using their outputs directly for evaluation or by including their outputs to train a second DNN.

For reverberant speaker separation, we additionally compare our system with a popular and representative time-domain algorithm, DPRNN-TasNet \cite{Luo2020}, which predicts the anechoic waveform from the noisy-reverberant one.

For noisy-reverberant speaker separation, we additionally compare our system with Wavesplit \cite{Zeghidour2020}, a state-of-the-art model in speaker separation. It is also a two-DNN system, but jointly trained, with the first one trained for computing speaker embeddings and the second performing speaker extraction based on the computed embeddings.
As a two-DNN system, the proposed method has some similarities to Wavesplit, but our system does not leverage any speaker embeddings.
Instead, we feed the separated speech obtained by $\text{DNN}_1$ and the convolutive-prediction results to $\text{DNN}_2$ for speaker extraction.
The encoder of $\text{DNN}_2$ may automatically encode some speaker information from the separated speech provided by $\text{DNN}_1$.

\ZQHL{The DPRNN-TasNet model contains around 3.6 million parameters, and the WaveSplit system contains around 29 million parameters according to \cite{Subakan2021}.
Each DNN model in our system contains around 6.9 million parameters.
Note that reducing the number of parameters is not a focus of this study.}

\section{Evaluation Results}\label{results}

\ZQHL{This section presents evaluation results on various tasks.
We also report results showing the effectiveness of FCP at reducing early reflections, and its capability at reverberation reduction.}

\subsection{Speech Dereverberation}\label{dereverbresults}

\begin{table*}[t]
\footnotesize
\centering
  \sisetup{table-format=2.2,round-mode=places,round-precision=2,table-number-alignment = center,detect-weight=true,detect-inline-weight=math}
\caption{\footnotesize{SI-SDR (dB), SDR (dB), PESQ, eSTOI (\%) and WER (\%) results on SMS-WSJ test set.}}
\label{resultssmswsj}
\begin{tabular}{lcccS[table-format=2.1,round-precision=1]S[table-format=2.1,round-precision=1]S[table-format=1.1]!{\enspace}S[table-format=2.1,round-precision=1]S}
\toprule %
Approaches & $\text{DNN}_1$ loss  & $\text{DNN}_2$ loss & $\text{DNN}_2$ type & \multicolumn{1}{c}{SI-SDR} & \multicolumn{1}{c}{SDR} & {PESQ}\hspace{-.5em}
& \hspace{-.2em}eSTOI & \multicolumn{1}{c}{WER} \\ %

\midrule

Unprocessed & - & - & - & -5.5 & -0.4 & 1.50 & 44.1 & 78.42 \\ %

\midrule %

$\text{DNN}_{1}$ & PIT & - & - & 5.6 & 7.657 & 2.06 & 72.1 & 42.64 \\ %
$\text{DNN}_{1}$ & PIT+sumPIT & - & - & 6.1 & 7.967 & 2.17 & 73.6 & 38.42 \\ %
\quad $\text{DNN}_{1}$+$\text{DNN}_{2}$ & PIT+sumPIT & RI & allSpks & 8.0 & 9.56886 & 2.25 & 77.3 & 36.67 \\ %
\quad $\text{DNN}_{1}$+$\text{DNN}_{2}$ & PIT+sumPIT & RI & perSpks & 9.8 & 11.2490 & 2.64 & 83.7 & 23.39 \\

\midrule %

\quad $\text{DNN}_{1}$+sfWPE+$\text{DNN}_2$ & PIT+sumPIT & RI & allSpks & 8.7 & 10.3517 & 2.38 & 80.1 & 31.38 \\ %
\quad $\text{DNN}_{1}$+sfWPE+$\text{DNN}_2$ & PIT+sumPIT & RI & perSpk & 11.2 & 12.62828 & 2.85 & 86.4 & 18.50 \\ %
\quad $\text{DNN}_{1}$+(sfWPE+$\text{DNN}_2$)$\times$2 & PIT+sumPIT & RI & perSpk & 11.5 & 12.9434 & 2.93 & 88.2 & 17.67 \\ %

\quad $\text{DNN}_{1}$+mfWPE+$\text{DNN}_2$ & PIT+sumPIT & RI & allSpks & 8.5 & 10.0558 & 2.35 & 79.1 & 32.40 \\ %
\quad $\text{DNN}_{1}$+mfWPE+$\text{DNN}_2$ & PIT+sumPIT & RI & perSpk & 11.0 & 12.41568 & 2.81 & 86.0 & 18.82 \\ %
\quad $\text{DNN}_{1}$+(mfWPE+$\text{DNN}_2$)$\times$2 & PIT+sumPIT & RI & perSpk & 11.4 & 12.78874 & 2.88 & 87.8 & 18.23 \\

\midrule %

\quad $\text{DNN}_{1}$+ICP+$\text{DNN}_2$  & PIT+sumPIT & RI & allSpks & 8.2 & 9.93675 & 2.30 & 78.0 & 34.49 \\ %
\quad $\text{DNN}_{1}$+ICP+$\text{DNN}_2$  & PIT+sumPIT & RI & perSpk & 10.7 & 12.1993 & 2.77 & 85.5 & 20.15 \\

\midrule %

\quad $\text{DNN}_{1}$+FCP+$\text{DNN}_2$ & PIT+sumPIT & RI & allSpks & 9.8 & 11.44447 & 2.53 & 81.8 & 27.79 \\ %
\quad $\text{DNN}_{1}$+FCP+$\text{DNN}_2$ & PIT+sumPIT & RI & perSpk & 12.0 & 13.43253 & 2.89 & 87.2 & 18.26 \\ %
\quad $\text{DNN}_{1}$+(FCP+$\text{DNN}_2$)$\times$2 & PIT+sumPIT & RI & perSpk & \bfseries 13.0 & \bfseries 14.41408 & 3.01 & 89.4 & 16.33 \\

\midrule %

\quad $\text{DNN}_{1}$+FCP+$\text{DNN}_2$ & PIT+sumPIT & RI+Mag & perSpk & 11.8 & 13.260519 & 3.22 & 88.1 & 13.53 \\ %
\quad $\text{DNN}_{1}$+(FCP+$\text{DNN}_2$)$\times$2 & PIT+sumPIT & RI+Mag & perSpk & 12.7 & 14.0847 & \bfseries 3.25 & \bfseries 89.9 & \bfseries 12.77 \\

\midrule %
\midrule

$\text{SISO}_1$ \cite{Wang2020c} & - & - & {-} & 5.1 & {-} & 2.44 & 74.6 & 28.28 \\ %
DPRNN-TasNet \cite{Luo2020} & - & - & - & 6.5 & {-} & 2.28 & 73.1 & 38.12 \\ %
6-microphone $\text{SISO}_1\text{-BF-}\text{SISO}_2$ \cite{Wang2020c} & - & - & - & 11.2 & {-} & 3.34 & 89.5 & 10.99 \\

\midrule %
Oracle direct sound + early reflections & - & - & - & {-} & {-} & {-} & {-} & 7.04 \\ %
Oracle spectral magnitude mask & - & - & - & 1.8 & 7.9 & 3.37 & 90.4 & 6.74 \\ %
Oracle phase-sensitive mask & - & - & - & 6.0 & 10.1 & 3.65 & 90.2 & 6.51 \\ %
Oracle direct sound & - & - & - & {-} & {-} & {-} & {-} & 6.40 \\

\bottomrule %
\end{tabular}
\vspace{-0.2cm}
\end{table*}

Table \ref{resultsdereverb} reports the results on our simulated test data for dereverberation, and on the ASR task of REVERB. %
In the literature \cite{Kinoshita2017, Haeb-Umbach2020, Drude2020}, the $\text{DNN}_1$ model in $\text{DNN}_1$-WPE can be trained to predict direct sound (denoted as ``d''), direct sound plus early reflections (denoted as ``d+e''), and direct sound plus early reflections and noise (denoted as ``d+e+v'').
In the case of our proposed convolutive prediction, $\text{DNN}_1$ is always trained to predict direct sound. For all the systems, $\text{DNN}_2$ is always trained to predict direct sound. 

Using direct sound (i.e., ``d'') as the training target for $\text{DNN}_1$ shows better performance over the other two (i.e., ``d+e'' and ``d+e+v''), if we consider the outputs of $\text{DNN}_1$ as the final prediction.
Comparing using different training targets for $\text{DNN}_1$, we do not observe a large performance difference in $\text{DNN}_1$-WPE, which applies $\text{DNN}_1$ outputs to improve WPE, but we notice that using direct sound to train $\text{DNN}_1$ shows the best performance in $\text{DNN}_1$+$\text{DNN}_2$, which stacks two DNNs by using the mixture and $\text{DNN}_1$ outputs as inputs to train $\text{DNN}_2$.
In the subsequent experiments, $\text{DNN}_1$ is always trained to estimate direct sound.

We then insert ICP, FCP, or WPE in between $\text{DNN}_1$ and $\text{DNN}_2$.
We first look at the effects of using different $\hat\lambda$ for linear prediction.
For the ICP in $\text{DNN}_1$+ICP+$\text{DNN}_2$, setting $\hat{\lambda}$ to Eq.~(\ref{complexproj1lambda}) and $\varepsilon$ to $1.0$ (i.e., no weighting) produces the best performance.
For the FCP in $\text{DNN}_1$+FCP+$\text{DNN}_2$, setting $\hat{\lambda}$ to Eq.~(\ref{complexproj2lambda}) and $\varepsilon$ to $0.001$ leads to the best performance.
For WPE, we can also compute $\hat\lambda$ based on Eq.~(\ref{complexproj2lambda}) with $\varepsilon$ set to $0.001$, meaning that we run the vanilla WPE for only one iteration (denoted as vWPE (1iter)).
$\text{DNN}_{1}$+vWPE (1iter)+$\text{DNN}_{2}$, which is essentially the same as $\text{DNN}_{1}$+$\text{DNN}_{2}$ but with $\text{DNN}_{2}$ additionally taking vWPE (1iter) results as features, shows worse performance than $\text{DNN}_1$+WPE+$\text{DNN}_2$, where $\hat\lambda$ is set to Eq.~(\ref{complexproj1lambda}) and $\varepsilon$ to $0.001$.
Among all setups for the linear predictions in between the two DNNs, $\text{DNN}_1$+FCP+$\text{DNN}_2$ with $\hat\lambda$ set to Eq.~(\ref{complexproj2lambda}) and $\varepsilon$ to $0.001$ shows the best performance.
By performing linear prediction and $\text{DNN}_2$ for one more iteration at run time, $\text{DNN}_{1}$+(WPE+$\text{DNN}_{2}$)$\times$2 and $\text{DNN}_{1}$+(ICP+$\text{DNN}_{2}$)$\times$2 show slight improvement in SI-SDR and PESQ and slight degradation in WER, while $\text{DNN}_{1}$+(FCP+$\text{DNN}_{2}$)$\times$2 shows clear improvements in all the metrics.
These results indicate the effectiveness of the proposed $\text{DNN}_{1}$+FCP+$\text{DNN}_{2}$ approaches over WPE and $\text{DNN}_{1}$+WPE+$\text{DNN}_{2}$.
\ZQHLAQ{They also indicate that, given the better phase and magnitude produced by the first-pass DNN$_2$ over DNN$_1$, FCP can better improve the second-pass DNN$_2$, while DNN-WPE only leverages the DNN-estimated magnitude as a weight in the objective function and the improvement in the second pass is relatively smaller.}

As we observe better scores in $\text{DNN}_1$+ICP+$\text{DNN}_2$ by setting $\hat\lambda$ to Eq.~(\ref{complexproj1lambda}) and $\varepsilon$ to $1.0$, while setting them to Eq.~(\ref{complexproj2lambda}) and $0.001$ leads to better scores in $\text{DNN}_1$+FCP+$\text{DNN}_2$, we use these respective setups for ICP and FCP in the following experiments.

Notice that if linear prediction results are used as the final outputs, $\text{DNN}_1$+FCP and $\text{DNN}_1$+ICP obtain better SI-SDR ($3.2$ and $3.6$ vs.\ $-1.0$ dB) and PESQ ($1.78$ and $1.82$ vs.\ $1.75$) than $\text{DNN}_1$+WPE, but worse WER ($18.55$\% and $17.55$\% vs.\ $15.02$\%).
The degradation in WER is possibly due to the fact that ICP and FCP do not have a prediction delay, while WPE does, so there could be more speech distortion.
\ZQHLAQ{
They also show worse SDR scores (5.9 and 4.5 vs. 6.4 dB), but some caution should be exercised when interpreting these results.
Indeed, this is possibly because the early reflections, which are not suppressed by $\text{DNN}_1$+WPE and could be well approximated by using a 512-tap filter (as allowed by SDR) to manipulate the direct-path signal, are considered as part of the target signal and could therefore boost the target energy when computing the SNR-like score.
On the other hand, FCP and ICP aim at reducing all reflections, but could introduce non-linear artifacts that cannot be approximated by linearly filtering the direct-path signal with a 512-tap filter.
See also our experiments in Section~\ref{reducingearlyresults}.
}

Overall, for speech dereverberation we improve SI-SDR and PESQ from the mixture's $-3.6$ dB and $1.64$ to $8.2$ dB and $2.65$ using a single-DNN system ($\text{DNN}_1$), to $9.1$ dB and $2.82$ using a plain two-DNN stacking system ($\text{DNN}_1$+$\text{DNN}_2$), to $12.3$ dB and $3.18$ by adding an FCP module in between the two DNNs ($\text{DNN}_{1}$+FCP+$\text{DNN}_2$), and to $13.3$ dB and $3.30$ by using one extra iteration for FCP and $\text{DNN}_2$ ($\text{DNN}_{1}$+(FCP+$\text{DNN}_{2}$)$\times$2).

We can add a magnitude domain loss during the training of $\text{DNN}_2$, following \cite{Wang2020a, Wang2020b}.
Clear improvements are obtained on WER and PESQ, while SI-SDR drops by around $0.6$ dB. This aligns with the findings in \cite{Wang2020a, Wang2020b}.

\ZQHLAQ{The 7.9\% WER obtained by our monaural system is better than the eight-microphone ``WPE+BeamformIt'' result, which comes with the Kaldi recipe for REVERB.}

\begin{table*}[t]
\footnotesize
\centering
  \sisetup{table-format=2.2,round-mode=places,round-precision=2,table-number-alignment = center,detect-weight=true,detect-inline-weight=math}
\caption{\footnotesize{SI-SDR (dB), SDR (dB), PESQ and eSTOI (\%) results on WHAMR! test set.}}
\label{resultswhamr!}
\begin{tabular}{lcccS[table-format=2.1,round-precision=1]S[table-format=2.1,round-precision=1]S[table-format=1.1]S[table-format=2.1,round-precision=1]}
\toprule %
Approaches & $\text{DNN}_1$ loss & $\text{DNN}_2$ loss & $\text{DNN}_2$ type & {SI-SDR} & {SDR} & {PESQ} & {eSTOI} \\

\midrule %
Unprocessed & - & - & - & -6.1 & -3.4865 & 1.41 & 31.7 \\ %

\midrule%

$\text{DNN}_{1}$ & PIT & RI & - & 2.9 &  5.1774 & 1.61 & 54.1 \\ %
$\text{DNN}_{1}$ & PIT+sumPIT & RI & - & 4.2 & 6.2283 & 1.79 & 59.4 \\ %
\quad $\text{DNN}_{1}$+$\text{DNN}_2$ & PIT+sumPIT & RI & allSpks & 5.6 & 7.0867 & 1.76 & 61.9 \\ %
\quad $\text{DNN}_{1}$+$\text{DNN}_2$ & PIT+sumPIT & RI & perSpks & 6.4 & 7.9414 & 1.93 & 68.5 \\

\midrule %

\quad $\text{DNN}_{1}$+sfWPE+$\text{DNN}_2$ & PIT+sumPIT & RI & allSpks & 5.8 & 7.34619 & 1.78 & 63.2 \\ %
\quad $\text{DNN}_{1}$+sfWPE+$\text{DNN}_2$ & PIT+sumPIT & RI & perSpk & 6.7 & 8.2409 & 1.95 & 69.4 \\ %
\quad $\text{DNN}_{1}$+(sfWPE+$\text{DNN}_2$)$\times$2 & PIT+sumPIT & RI & perSpk & 6.4 & 7.929 & 1.96 & 71.5 \\ %

\quad $\text{DNN}_{1}$+mfWPE+$\text{DNN}_2$ & PIT+sumPIT & RI & allSpks & 5.8 & 7.2465 & 1.79 & 63.0 \\ %
\quad $\text{DNN}_{1}$+mfWPE+$\text{DNN}_2$ & PIT+sumPIT & RI & perSpk & 6.8 & 8.28229 & 1.96 & 69.5 \\ %
\quad $\text{DNN}_{1}$+(mfWPE+$\text{DNN}_2$)$\times$2 & PIT+sumPIT & RI & perSpk & 6.6 & 7.979 & 1.96 & 71.7 \\

\midrule %

\quad $\text{DNN}_{1}$+ICP+$\text{DNN}_2$ & PIT+sumPIT & RI & allSpks & 5.8 & 7.45898 & 1.82 & 63.4 \\ %
\quad $\text{DNN}_{1}$+ICP+$\text{DNN}_2$ & PIT+sumPIT & RI & perSpk & 6.7 & 8.2194 & 1.95 & 69.3 \\

\midrule %

\quad $\text{DNN}_{1}$+FCP+$\text{DNN}_2$ & PIT+sumPIT & RI & allSpks & 6.4 & 7.899 & 1.83 & 64.3 \\ %
\quad $\text{DNN}_{1}$+FCP+$\text{DNN}_2$ & PIT+sumPIT & RI & perSpk & 7.4 & 8.79691 & 1.97 & 70.1 \\ %
\quad $\text{DNN}_{1}$+(FCP+$\text{DNN}_2$)$\times$2 & PIT+sumPIT & RI & perSpk & \bfseries 7.5 & \bfseries 8.977398 & 2.01 & 72.7  \\

\midrule %

\quad $\text{DNN}_{1}$+FCP+$\text{DNN}_2$ & PIT+sumPIT & RI+Mag & perSpk & 7.3 & 8.75159 & \bfseries 2.39 & 72.2 \\ %
\quad $\text{DNN}_{1}$+(FCP+$\text{DNN}_2$)$\times$2 & PIT+sumPIT & RI+Mag & perSpk & 7.4 & 8.8532 & \bfseries 2.39 & \bfseries 74.3 \\

\midrule\midrule
Conv-TasNet \cite{Luo2019, Maciejewski2020} & - & - & - & 2.2 & {-} & {-} & {-} \\ %
3-Stage BLSTM-TasNet \cite{Maciejewski2020} & - & - & - & 4.8 & {-} & {-} & {-} \\ %
Wavesplit \cite{Zeghidour2020} & - & - & - & 5.9 & {-} & {-} & {-} \\ %
\bottomrule %
\end{tabular}
\vspace{-0.2cm}
\end{table*}

\subsection{Reverberant Speaker Separation}

Table \ref{resultssmswsj} reports the performance on SMS-WSJ as well as oracle results obtained by using \ZQHL{as estimate} the direct sound with or without early reflections and oracle masks such as the spectral magnitude mask ($|S|/|Y|$) \cite{WYX2014} and phase-sensitive mask ($|S|/|Y|\text{cos}(\angle S-\angle Y)$) \cite{Erdogan2015}.
Notice that using oracle direct sound for ASR obtains better WER over using direct sound plus early reflections ($6.40$\% vs.\ $7.04$\%).
This indicates the potential benefits of removing early reflections, \ZQHLAQ{which can slightly smear spectral patterns.}

Compared to training $\text{DNN}_1$ with $\mathcal{L}^{(1)}_{\text{PIT}}$, using $\mathcal{L}^{(1)}_{\text{PIT+sumPIT}}$ for training shows better performance.
The plain two-DNN stacking system, $\text{DNN}_1$+$\text{DNN}_2$, shows consistent improvements over $\text{DNN}_1$.

For DNN-WPE, we explore two variants for multi-speaker scenarios.
The first one uses the PSD of each estimated target speaker produced by $\text{DNN}_{1}$ to compute a different WPE filter for each speaker. We denote this algorithm as $\text{DNN}_{1}$+mfWPE+$\text{DNN}_2$, where ``mf'' means multi-filter.
The other one, following \cite{Zhang2020,Haeb-Umbach2020}, sums up all the estimated target speakers provided by $\text{DNN}_{1}$ and uses the PSD of the summated signal to compute a single WPE filter to dereverberate the mixture.
We denote this variant as $\text{DNN}_{1}$+sfWPE+$\text{DNN}_2$, where ``sf'' means single-filter.
From Table \ref{resultssmswsj}, we notice that $\text{DNN}_{1}$+sfWPE+$\text{DNN}_2$ obtains slightly better performance than $\text{DNN}_{1}$+mfWPE+$\text{DNN}_2$.
This suggests that computing a separate filter for each target speaker does not \ZQHLAQ{bring improvements over DNN-WPE that uses a single filter for all speakers}.

Let us first consider the case (denoted as ``allSpks'' in Table \ref{resultssmswsj}) where $\text{DNN}_2$ is trained to enhance all the target speakers altogether as in the system in Fig.~\ref{systemspksep1}. Compared with $\text{DNN}_{1}$+sfWPE+$\text{DNN}_2$ and $\text{DNN}_{1}$+ICP+$\text{DNN}_2$, $\text{DNN}_{1}$+FCP+$\text{DNN}_2$ shows better performance in all the metrics.
This demonstrates the effectiveness of FCP over WPE at dereverberation when competing speakers are present.
If we instead train $\text{DNN}_2$ to enhance target speakers one by one as in the system in Fig.~\ref{systemspksep2} (denoted as ``perSpk'' in Table \ref{resultssmswsj}), we obtain further improvement.
This suggests that dereverberating each speaker individually helps. Further iterating convolutive prediction and $\text{DNN}_2$ for one more iteration leads to consistent improvement. Again, training $\text{DNN}_2$ by including a magnitude-level loss as in \cite{Wang2020a, Wang2020b} improves PESQ, eSTOI, and WER, but slightly decreases SI-SDR.

Our best performing system obtains much better results over $\text{SISO}_1$, another complex spectral mapping system recently proposed in \cite{Wang2020c} ($13.0$ vs.\ $5.1$ dB SI-SDR), and over DPRNN-TasNet \cite{Luo2020} ($13.0$ vs.\ $6.5$ dB SI-SDR).
Surprisingly, our best single-channel system is even comparable to a strong six-microphone system, $\text{SISO}_1\text{-BF-}\text{SISO}_2$ proposed recently in \cite{Wang2020c}, which combines monaural complex spectral mapping with beamforming and post-filtering.
\ZQHL{These results suggest that combining DNNs operating in the complex domain with convolutive prediction
is very effective at reverberation suppression}, and further integrating it with multi-microphone processing is a promising direction for future research.

\subsection{Noisy-Reverberant Speaker Separation}

Table \ref{resultswhamr!} reports the separation results on WHAMR!.
A similar trend as in Table \ref{resultssmswsj} is observed. $\text{DNN}_{1}$+FCP+$\text{DNN}_2$ produces better results over  $\text{DNN}_{1}$+mfWPE+$\text{DNN}_2$ ($7.4$ vs.\ $6.8$ dB SI-SDR). This indicates that DNN-FCP is more robust than DNN-WPE at dereverberation when noises and competing speakers are present.

\begin{table}[t]
\footnotesize
\scriptsize
  \sisetup{table-format=2.2,round-mode=places,round-precision=2,table-number-alignment = center,detect-weight=true,detect-inline-weight=math}
\caption{\footnotesize{SI-SDR (dB), SDR (dB), and PESQ performance of DNN-FCP and DNN-WPE for reducing early reflections on \ZQHLAQ{Dereverb Data II}.
}}
\label{reducingearly}
\setlength{\tabcolsep}{4.5pt}
\begin{tabular}{lcccS[table-format=2.1,round-precision=1]S[table-format=2.1,round-precision=1]S[table-format=1.1]}
\toprule %
Approaches & DNN$_1$ predicts? & $\hat{\lambda}$ & $K$\,/\,$\Delta$\,/\,$\varepsilon$ & {SI-SDR} & {SDR} & {PESQ} \\
\midrule %
Unprocessed & - & - & - & -1.4 & 7.88924 & 2.26 \\ 

\midrule %

$\text{DNN}_1$-WPE & d & (\ref{complexproj1lambda}) & 39\,/\,1\,/\,$0.001$ & 1.3 & 12.070157 & 2.51 \\ %
$\text{DNN}_1$-WPE & d & (\ref{complexproj1lambda}) & 38\,/\,2\,/\,$0.001$ & 2.0 & \bfseries 12.4537 & 2.46 \\ %
$\text{DNN}_1$-WPE & d & (\ref{complexproj1lambda}) & 37\,/\,3\,/\,$0.001$ & 1.5 & 11.584572 & 2.42 \\ %
$\text{DNN}_1$-WPE & d & (\ref{complexproj1lambda}) & 36\,/\,4\,/\,$0.001$ & 0.6 & 10.893 & 2.38 \\ %
$\text{DNN}_1$-FCP & d & (\ref{complexproj2lambda}) & 40\,/\,0\,/\,$0.001$ & \bfseries 8.2 & 10.8356 & \bfseries 2.96 \\
\bottomrule %
\end{tabular}
\end{table}

Compared with Wavesplit \cite{Zeghidour2020}, which reports the best results to date on WHAMR!, our system obtains clearly better SI-SDR ($7.5$ vs.\ $5.9$ dB).
Wavesplit uses speaker identities as a side information during training for target speaker extraction, while our system does not rely on the availability of such information.
Wavesplit also considers applying dynamic mixing for data augmentation, leading to better SI-SDR ($7.1$ dB) \cite{Zeghidour2020}.
Even without such data augmentation, our system still obtains a better result than Wavesplit's result with dynamic mixing.
\remove{Future work will consider adding dynamic mixing to our training setup, allowing for same-speaker mixtures so as not to use speaker identities as side information.}

\subsection{FCP's Effectiveness at Reducing Early Reflections}\label{reducingearlyresults}

Since DNN-FCP does not require a prediction delay and can leverage both magnitude and phase estimated by DNNs for filter estimation, it has the potential to better reduce early reflections than DNN-WPE.
To support this claim, we design an experiment based on the speech dereverberation task.
For each mixture in the test set of Dereverb Data I, we remove the late reverberation and the stationary air-conditioning noise, each mixture thus only containing the direct-path and early reflections (See Eq.~(\ref{eq:phymodel_time}) for the definition of early reflections and late reverberation).
We denote this new dataset as ``\ZQHLAQ{Dereverb Data II}''.
We then feed the new mixtures directly to the well-trained $\text{DNN}_1$ (presented in Table~\ref{resultsdereverb}), and compare the SI-SDR scores of DNN-WPE and DNN-FCP.
Note that the direct-path signal is used as the reference signal for computing SI-SDR, and hence the system with the higher SI-SDR indicates that it is better at \ZQHLAQ{predicting the direct-path signal and} suppressing early reflections.
\ZQHLAQ{As shown in Table~\ref{reducingearly}, DNN-FCP gets clearly better SI-SDR, as well as PESQ.
The SDR result is slightly lower, possibly because the 512-tap filter used internally by SDR could not compensate for non-linear artifacts introduced by FCP as discussed in the fifth paragraph of Section~\ref{dereverbresults} for a related experiment.
}

\ZQHLAQ{To verify that the low SI-SDR scores of WPE are not due to WPE silently introducing a small time-invariant signal shift in its processing results, 
we used the GCC-PHAT algorithm \cite{DiBiase2001}, a popular algorithm for time difference of arrival estimation, to time-align the WPE result with the direct-path reference signal before computing SI-SDR.
Our finding is that the best time delay computed by GCC-PHAT is always zero for all the utterances in the test set, suggesting that the low SI-SDR scores are not because WPE incurs a signal shift.}

\section{Conclusion}\label{conclusion}

We have proposed a convolutive prediction approach for reverberation suppression.
Evaluation results on speech dereverberation and speaker separation show the effectiveness of the proposed algorithm over the popular DNN-WPE algorithm.
In addition, our study delivers a message that although plain end-to-end modeling based on advanced neural network architectures is effective at suppressing reverberation, combining it with techniques based on domain knowledge, for example with the proposed convolutive prediction or the DNN-WPE algorithm, leads to large improvements.
Although our DNN is a TCN-DenseUNet model trained in the complex domain, it can be readily replaced by magnitude- or time-domain models and by more advanced DNN architectures.
In other words, our algorithms can ride on the development of end-to-end neural networks, as better anechoic target speech estimated by a DNN is likely to lead to better convolutive prediction.

Similar to DNN-WPE, the proposed convolutive prediction has closed-form solutions.
This makes our algorithm suitable for online real-time processing and capable of being jointly trained with other DNN modules such as acoustic models.
Our future work shall backpropagate through convolutive prediction and train all DNNs end-to-end.
In addition, we will extend the proposed algorithms to multi-microphone scenarios and evaluate them on real recordings such as LibriCSS \cite{Chen2020LibriCSS} and CHiME-5 \cite{Barker2018CHiME5}.

In closing, we emphasize that the linear-filter structure in reverberation provides an informative cue for dereverberation, and explicitly leveraging it using deep learning supported convolutive prediction could be an important step towards solving the cocktail party problem in realistic conditions.

\bibliographystyle{IEEEtran}
\bibliography{references.bib}

\vspace{-33pt}

\begin{IEEEbiography}[{\includegraphics[width=1in,height=1.25in,clip,keepaspectratio]{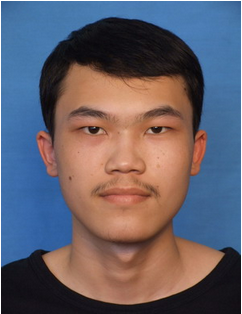}}]{Zhong-Qiu Wang} is a Postdoctoral Research Associate at Carnegie Mellon University, PA, USA. He received the B.E. degree in computer science and technology from Harbin Institute of Technology, Harbin, China, in 2013, and the M.S. and Ph.D. degrees in computer science and engineering from The Ohio State University, Columbus, OH, USA, in 2017 and 2020, respectively. His research interests include microphone array processing, speech separation, robust automatic speech recognition, machine learning, and deep learning.
\end{IEEEbiography}

\vspace{11pt}

\begin{IEEEbiography}
[{\includegraphics[width=1in,height=1.25in,clip,keepaspectratio]{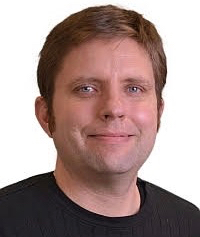}}]{Gordon Wichern}
is a Principal Research Scientist at Mitsubishi Electric Research Laboratories (MERL) in Cambridge, Massachusetts. He received his B.Sc.\ and M.Sc.\ degrees from Colorado State University in electrical engineering and his Ph.D.\ from Arizona State University in electrical engineering with a concentration in arts, media and engineering, where he was supported by a National Science Foundation (NSF) Integrative Graduate Education and Research Traineeship (IGERT) for his work on environmental sound recognition.  He was previously a member of the research team at iZotope, inc.\ where he focused on applying novel signal processing and machine learning techniques to music and post production software, and a member of the Technical Staff at MIT Lincoln Laboratory where he worked on radar signal processing.  His research interests include audio, music, and speech signal processing, machine learning, and psychoacoustics.
\end{IEEEbiography}

\vspace{11pt}

\begin{IEEEbiography}[{\includegraphics[width=1in,height=1.25in,clip,keepaspectratio]{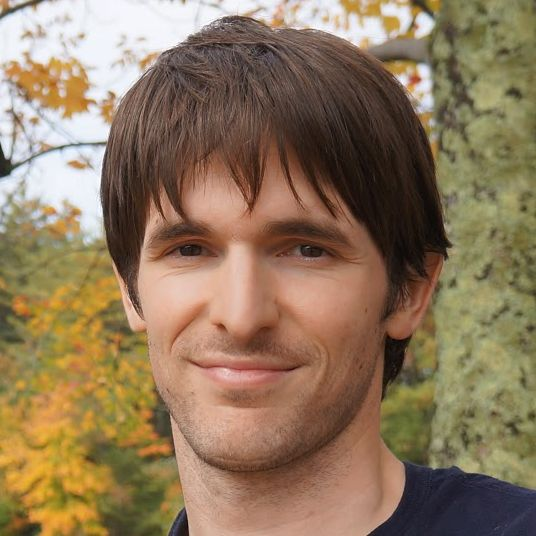}}]{Jonathan Le Roux} is a Senior Principal Research Scientist and the Speech and Audio Senior Team Leader at Mitsubishi Electric Research Laboratories (MERL) in Cambridge, Massachusetts. He completed his B.Sc.\ and M.Sc.\ degrees in Mathematics at the Ecole Normale Sup\'erieure (Paris, France), his Ph.D.\ degree at the University of Tokyo (Japan) and the Universit\'e Pierre et Marie Curie (Paris, France), and worked as a postdoctoral researcher at NTT’s Communication Science Laboratories from 2009 to 2011. His research interests are in signal processing and machine learning applied to speech and audio. He has contributed to more than 100 peer-reviewed papers and 20 granted patents in these fields. He is a founder and chair of the Speech and Audio in the Northeast (SANE) series of workshops, and a Senior Member of the IEEE.
\end{IEEEbiography}

\end{document}